\documentclass[twocolumn,superscriptaddress]{revtex4-1}
\usepackage{lmodern}
\usepackage[utf8x]{inputenc}
\usepackage{amsmath}
\usepackage{graphicx}
\usepackage{amsfonts}
\usepackage{epstopdf}
\usepackage{placeins}
 \usepackage{verbatim}
\usepackage{bbold}

\usepackage{url}
\usepackage{epstopdf}
\usepackage{hyperref}
\hypersetup{breaklinks=true}

\usepackage{xcolor}

\begin{document}

\title{Long-lived elementary excitations and light coherence in topological lasers}

\author{Petr Zapletal}

\affiliation{Cavendish Laboratory, University of Cambridge, Cambridge CB3 0HE, United Kingdom}

\author{Bogdan Galilo}

\affiliation{Cavendish Laboratory, University of Cambridge, Cambridge CB3 0HE, United Kingdom}

\author{Andreas Nunnenkamp}

\affiliation{Cavendish Laboratory, University of Cambridge, Cambridge CB3 0HE, United Kingdom}

\affiliation{School of Physics and Astronomy and Centre for the Mathematics and Theoretical Physics of Quantum Non-Equilibrium Systems, University of Nottingham, Nottingham, NG7 2RD, UK}

\date{\today}

\begin{abstract}
Combining topologically-protected chiral light transport and laser amplification gives rise to topological lasers, whose operation is immune to fabrication imperfections and defects, uncovering the role of topology in a novel nonlinear non-Hermitian regime. We study a topological laser based on the photonic Haldane model with selective pumping of chiral edge modes described by saturable gain. We investigate elementary excitations around the mean-field steady state and their consequences for the coherence properties. In particular, we show that the hybridization of chiral edge modes gives rise to long-lived elementary excitations, leading to large phase fluctuations in the emitted light field and a decrease of light coherence. In contrast to topologically trivial lasers, the lifetime of elementary excitations is robust against disorder in topological lasers. However, the lifetime strongly depends on the edge-mode dispersion around the lasing frequency. As a result, the lifetime can be reduced by orders of magnitude for lasing of different edge modes, leading to a suppression of phase fluctuations and larger coherence of the emitted light. On the other hand, amplitude fluctuations and the second-order autocorrelation function are moderately increased at the same time. 
\end{abstract}

\maketitle

\section{Introduction}
 
Topological photonics has made rapid strides in the past years \cite{ozawa2019}, investigating effects of gain and loss on the topology of photonic energy bands \cite{martinezalvarez2018,zhao2019,cerjan2019}, topology in synthetic dimensions \cite{yuan2016,ozawa2016,lustig2019,dutt2020}as well as the interplay of topology and nonlinear optics phenomena \cite{smirnova2020}. Lasing in topological photonic structures has recently attracted a lot of attention not only because it allows studying topology in a novel nonlinear non-Hermitian regime but also because topological structures can offer a new design of laser devices. First, lasing of zero-dimensional edge modes has been demonstrated in one-dimensional photonic arrays \cite{st-jean2017,parto2018,zhao2018}. These pioneering works have been followed by experiments reporting lasing of one-dimensional chiral edge modes in two-dimensional photonic arrays \cite{bahari2017b,bandres2018,klembt2018,zeng2020}. In a two-dimensional array, lasing of a single edge mode extending over the whole edge of the photonic array has been demonstrated \cite{bandres2018}. The single-mode laser operation is robust against on-site disorder in contrast to topologically trivial laser arrays \cite{harari2018}. For this reason, topological lasers are a promising candidate for highly-efficient lasers with a robust emission spectrum. The rich dynamics of topological lasers are subject to current theoretical investigation \cite{longhi2018,secli2019}. However, the theory for coherence properties of topological lasers, which would be relevant for recent experiments demonstrating stable laser operation \cite{bandres2018}, has been still missing. 

One essential characteristic of lasers is their large temporal coherence of the emitted light field, which is required for practical applications \cite{haken1985}. The coherence is fundamentally limited by the phase diffusion of the light field caused by the intrinsic noise due to spontaneous emission \cite{gardiner2004}. Phase diffusion leads to a finite linewidth of the emitted light field, which, in the absence of other noise sources, is determined by the Schawlow-Townes formula \cite{schawlow1958}.
In realistic lasers, the coherence of the emitted light is affected by the dynamics of gain medium as well as the presence of multiple lasing modes, leading to an additional broadening of the laser linewidth \cite{pick2015}.
In this manuscript, we study how the coherence of the light field emitted by a topological laser is affected by the elementary excitations around the mean-field steady state, which are excited by intrinsic noise. To focus on the effects of the elementary excitations, we neglect the dynamics of the gain medium assuming the gain medium instantaneously responds to the dynamics of the light field.
 
\begin{figure*}[htbp]
\centering
\includegraphics[width=\linewidth]{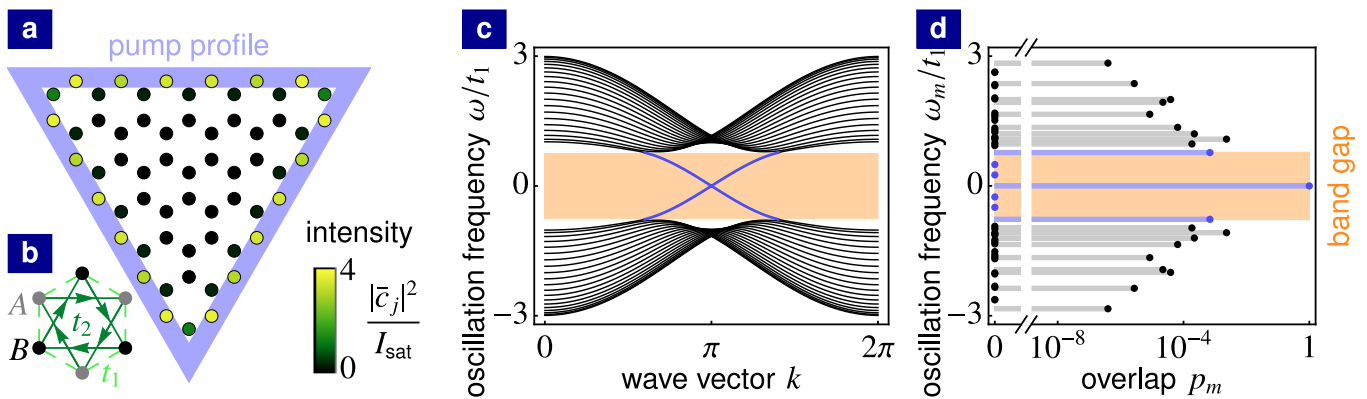}
\caption{Mean-field steady state of topological laser. (a) The honeycomb photonic array with photon tunneling described by the Haldane model pumped in the blue region. The color scale shows the mean-field steady state occupations $|\bar{c}_j|^2$ of local optical sites. (b) Unit cell of the Haldane model consisting of the sublattice $A$ (gray points), the sublattice $B$ (black points), the nearest-neighbor hopping with a real tunneling $t_1$ and the next-nearest-neighbor hopping with a complex amplitude $t_2 e^{\phi_{jk}}$. (c) Band structure of the passive Haldane model (no gain no loss) with bulk modes (black lines), topological band gap (orange region) and topological edge modes (blue lines) for an infinite strip with zig-zag edges. (d) The overlap $p_m$ of the mean-field steady-state solution with normal modes $\mathbf{e}^{(m)}$ of the passive system (no gain no loss) for the mode with frequency $\Omega/t_1=0$ lasing. (Parameters: $t_2/t_1=0.15$, $\phi=\pi/2$, $\gamma/t_1=0.01$, $g/t_1=0.05$, $N=61$)}
\label{steady}
\end{figure*}

We consider the Haldane model based on a two-dimensional photonic array pumped along the edge. On the mean-field level neglecting quantum and thermal fluctuations in the laser, we obtain lasing of a single edge mode. Depending on initial conditions, lasing of edge modes with different lasing frequencies can be achieved as it was described in Ref.~\cite{secli2019}. We take fluctuations into account using nonlinear semiclassical Langevin equations. We linearize the Langevin equations around the mean-field steady-state solution to study elementary excitations. We consider weak gain and loss in comparison to the coupling of optical sites in the array and a moderate size of the array such that the frequency separation of edge modes is larger that the linewidth of these modes. This regime is relevant for recent experiments on arrays of micron-scale ring resonators \cite{hafezi2013,bandres2018}. We study how normal modes of elementary excitations are formed from the normal modes of a passive system, which does not experience either gain or loss. We show that the hybridization of edge modes gives rise to long-lived elementary excitations, which lead to large phase fluctuations and a decreased coherence of the emitted light field. 
The emergence of long-lived elementary excitations is not a unique feature of topological lasers as they generically appear in laser arrays with a linear frequency dispersion.
However, the fact that the long-lived elementary excitations in a topological laser are formed from topological edge modes makes them robust against disorder. We show that, in contrast to long-lived elementary excitations in a trivial laser, the life-time and the oscillation frequency of these topological long-lived elementary excitations are robust against moderate on-site disorder.
The lifetime of elementary excitations strongly depends on the dispersion of edge-mode frequencies around the lasing frequency. Any deviation from a linear dispersion leads to a detuning for normal modes of elementary excitations, which can obstruct their hybridization and, as a consequence, reduce their lifetime. For lasing at frequencies, which do not lie in the middle of the passive-system band gap, the deviation from a linear dispersion is sufficient to reduce the lifetime of elementary excitations by at least one order of magnitude. This leads to a large suppression of phase fluctuations and an increase of light coherence. On the other hand, amplitude fluctuations of the emitted light field are increased resulting in a moderately larger second-order autocorrelation function. We confirm our results by numerical simulations of full Langevin equations, which take nonlinear noise dynamics into account.

\section{Model}

We consider an array of optical sites, whose complex amplitudes $c_j$, $j=1,...,N$, are described by the semiclassical Langevin equations
\begin{equation}\label{full eq}
i \frac{\rm d}{{\rm d} t}c_j = \left[ \nu_j  - i \gamma + i \frac{\mathbb{P}_j g}{1 + \frac{|c_j|^2}{I_{\rm sat}}} \right]c_j  + \sum^{N}_{k=1} H_{jk} c_k + Q_{jj} c_{j,{\rm in}},
\end{equation}
where $\hbar=1$, $\nu_j$ are the frequencies of the optical sites, the Hamiltonian $H_{jk}$ describes the coupling of these sites, and $N$ is the number of the optical sites in the array. Intrinsic optical losses lead to a decay at rate $\gamma$. Incoherent pumping of optical sites is described by a saturable gain $g$, where $I_{\rm sat}$ is the saturation intensity. We allow for a spatial pump profile where $\mathbb{P}_j = 1$ for pumped sites and $\mathbb{P}_j = 0$ for not pumped sites. 
Incoherent pumping is associated with intrinsic noise due to spontaneous emission at rate $q$, which is the dominant source of fluctuations at the pumped sites. At sites without pumping, the dominant source of fluctuations is shot noise at rate $2\gamma$. Both intrinsic noise due to spontaneous emission and shot noise can be described by Gaussian white noise $\langle c_{j,{\rm in}}(t)c^*_{k,{\rm in}}(t')\rangle = \delta_{jk} \delta(t-t')$ with a correlation matrix $\mathbf{Q}\mathbf{Q}^{\dagger}$, where $\mathbf{Q}$ is a diagonal matrix, $Q_{jk} = \delta_{jk}\left[\sqrt{2\gamma}(1-\mathbb{P}_j) +  \sqrt{q}\mathbb{P}_j\right]$, and $\delta_{jk}$ is the Kronecker delta.

We focus on the Haldane Hamiltonian $\hat{H} = t_1\sum_{\rm n.n.} \hat{c}^{\dagger}_j \hat{c}_k + t_2\sum_{\rm n.n.n.} e^{i\phi_{jk}} \hat{c}^{\dagger}_j \hat{c}_k$ based on a honeycomb array (see Fig~\ref{steady}b) including the nearest-neighbor hopping with a real amplitude $t_1$ and the next-nearest-neighbor hopping with a complex amplitude $t_2 e^{i\phi_{jk}}$ \cite{haldane1988,jotzu2014}. $\phi_{jk}=\phi$ for hopping in the directions shown by green arrows in Fig~\ref{steady}b and $\phi_{jk}=-\phi$ in the reverse directions, where $\phi$ is the Haldane flux.  In Fig.~\ref{steady}c, we plot the band structure of the passive Haldane model (black lines) for no gain and no loss in the photonic array. 
For $\phi \neq0,\pi$, the time-reversal symmetry of the system is broken and a topological band gap opens (orange region). Cutting the array in a form of an infinite strip, chiral edge modes (blue lines) appear at the boundaries of the array. Frequencies of the chiral edge modes lie in the topological band gap. 

\section{Mean-field steady state}

We first find steady states of the mean-field dynamical equations for optical amplitudes, which are obtained by omitting stochastic terms in the Langevin equations (\ref{full eq}). We consider a finite array depicted in Fig.~\ref{steady}a, where optical sites in the blue region are pumped. We assume that gain and loss are weak in comparison to the hopping amplitudes, i.e.~$g,\gamma\ll t_1,t_2$.  In this regime, lasing of a single topological edge mode is achieved, which was theoretically shown in Ref.~\cite{harari2018} and experimentally demonstrated in Ref.~\cite{bandres2018}. In Fig.~\ref{steady}d, we show the overlap $p_m=|\sum_{j=1}^N\bar{c}_j^*e_j^{(m)}|/\sqrt{\sum_{j=1}^N |\bar{c}_j|^2}$ of the mean-field steady state solution $\bar{c}_j$ with the normal modes $\mathbf{e}^{(m)}$ of the passive system (no gain no loss). Depending on the initial conditions, one of the edge modes wins the gain competition. Since all edge modes extend across the whole pump region, a single edge mode saturates gain at all pumped optical sites and prevents lasing of other edge modes. As a result, the overlap of the mean-field steady state with a single edge mode is close to unity and the overlaps with the remaining passive-system normal modes is very small. Lasing of different edge modes leads to different lasing frequencies and different steady-state distributions of optical phases $\bar{\theta}_j$ along the edge of the array (see Fig.~\ref{excitations}a and \ref{excitations}b). However, the occupation of optical sites, $|\bar{c}_j|^2$, (see Fig.~\ref{steady}a) is almost identical for lasing of any edge mode, since all edge modes have very similar spatial profile $|e_j^{(m)}|^2$. The mean-field dynamics of complex amplitudes $c_j/\sqrt{I_{\rm sat}}$ and the mean-field steady state $\bar{c}_j/\sqrt{I_{\rm sat}}$ are independent of the absolute scaling $I_{\rm sat}$.

\section{Elementary excitations}\label{sec elm}
In this section, we describe elementary excitations around the mean-field steady state. We show how the normal modes of elementary excitations are formed from the passive-system normal modes.

To study elementary excitations around the mean-field steady state, we decompose optical amplitudes $c_j = \left(\bar{c}_j + \delta c_j \right)e^{-i\Omega t}$ into the mean-field steady-state solution $\bar{c}_j $ and a modulation $\delta c_j$, where $\Omega$ is the frequency of the lasing mode. Considering small modulations around the mean-field steady state, we derive linear Langevin equations 
\begin{equation}\label{bog eq}
i\frac{\rm d}{{\rm d} t}
\begin{pmatrix}
\delta\mathbf{ c} \\
\delta\mathbf{ c}^*
\end{pmatrix}
=
\mathbf{\mathcal{D}}
\begin{pmatrix}
\delta \textbf{c} \\
\delta \textbf{c}^*
\end{pmatrix}
+
\mathcal{Q}
\begin{pmatrix}
\mathbf{ c}_{\rm in}e^{i\Omega t}  \\
\mathbf{ c}^*_{\rm in}e^{-i\Omega t} 
\end{pmatrix},
\end{equation}
where $\mathcal{D}$ is the dynamical matrix for elementary excitations around the mean-field steady state, $\mathcal{Q} = \mathbf{Q}\otimes\sigma_z$ and $\sigma_z$ is the Pauli matrix.
The dynamical matrix 
\begin{equation}\label{dyn m}
\mathcal{D} = \mathcal{H} + \mathcal{A}=
\begin{pmatrix}
\mathbf{H} - \Omega\,\mathbb{1} & 0 \\
0 & -\mathbf{H}^*  + \Omega\,\mathbb{1}
\end{pmatrix}
+
i\begin{pmatrix}
 \mathbf{\Gamma}  & \mathbf{\Delta} \\
\mathbf{\Delta}^* &  \mathbf{\Gamma}
\end{pmatrix}
\end{equation}
can be decomposed into the Hermitian part $\mathcal{H}$ and the anti-Hermitian part $\mathcal{A}$, where $H_{jk}$ is the Hamiltonian of the passive system,
$\mathbb{1}$ is the $N\times N$ identity matrix,
\begin{equation}
\Gamma_{jj} =  -\gamma + \frac{\mathbb{P}_{j}\,g}{\left(1 + \frac{|\bar{c}_{j}|^2}{I_{\rm sat}}\right)^2},\,\,\,\,\Delta_{jj} =  -  \frac{\mathbb{P}_{j}\,g\,\frac{\bar{c}_{j}^2}{I_{\rm sat}}}{\left(1 + \frac{ |\bar{c}_{j}|^2}{I_{\rm sat}}\right)^2},
\end{equation}
and $\Gamma_{jk}=\Delta_{jk}=0$ for $j\neq k$. The dynamical matrix depends only on rescaled mean-field optical amplitudes $\bar{c}_j/\sqrt{I_{\rm sat}}$. As a result, elementary excitations do not depend on the absolute scaling, $I_{\rm sat}$, of the mean-field optical amplitudes.

For elementary excitations, the number of normal modes is doubled compared to the number of the passive-system normal modes.  The dynamical matrix $\mathcal{D}$ exhibits the following symmetry
$\mathcal{X}\mathcal{D}\mathcal{X}=-\mathcal{D}^*$, where $\mathcal{X} = \mathbb{1}\otimes\sigma_x $ and $\sigma_x$ is the Pauli matrix. Due to this symmetry, the complex frequencies $\epsilon^{(\alpha)}$, $\alpha = 1,...,2N$, of elementary excitations are purely imaginary or appear in pairs $\left(\epsilon^{(\alpha)}, \tilde{\epsilon}^{(\alpha)}\right)$, where $\tilde{\epsilon}^{(\alpha)}= - \left(\epsilon^{(\alpha)}\right)^*$. 

We first diagonalize the Hermitian part $\mathcal{H}$ by switching to the basis of passive-system normal modes $\mathcal{E}_p^{(m)} = \mathbf{e}^{(m)}\otimes\left(1,0\right)^{T}$ and $\tilde{\mathcal{E}}_p^{(m)} = \left(\mathbf{e}^{(m)}\right)^*\otimes\left(0,1\right)^{T}$, where $\mathbf{e}^{(m)}$, $m=1,...,N$, are eigenmodes of $\mathbf{H}$. The eigenfrequencies of the Hermitian part are directly formed from the passive-system eigenfrequencies $\omega_m$ , giving rise to two branches $\epsilon_p^{(m)}=\omega_m - \Omega$ and $\tilde{\epsilon}_p^{(m)} = - \omega_m + \Omega$. The anti-Hermitian part 
\begin{equation}
\tilde{\mathcal{A}}  = 
i\begin{pmatrix}
 \tilde{\mathbf{\Gamma}}  & \tilde{\mathbf{\Delta}} \\
\tilde{\mathbf{\Delta}}^* &  \tilde{\mathbf{\Gamma}}^*
\end{pmatrix}
\end{equation}
introduces coupling between passive-system normal modes, where $\tilde{\mathbf{\Gamma}} = \mathbf{U}^{\dagger}\, \mathbf{\Gamma}\,\mathbf{U}$, $\tilde{\mathbf{\Delta}} = \mathbf{U}^{\dagger}\, \mathbf{\Delta}\,\mathbf{U}^{*}$. Columns of the transformation matrix $\mathbf{U}$ are eigenmodes $\mathbf{e}^{(m)}$. Due to the anti-Hermitian coupling, the passive-system normal modes hybridize.

\begin{figure*}[htbp]
\includegraphics[width=\linewidth]{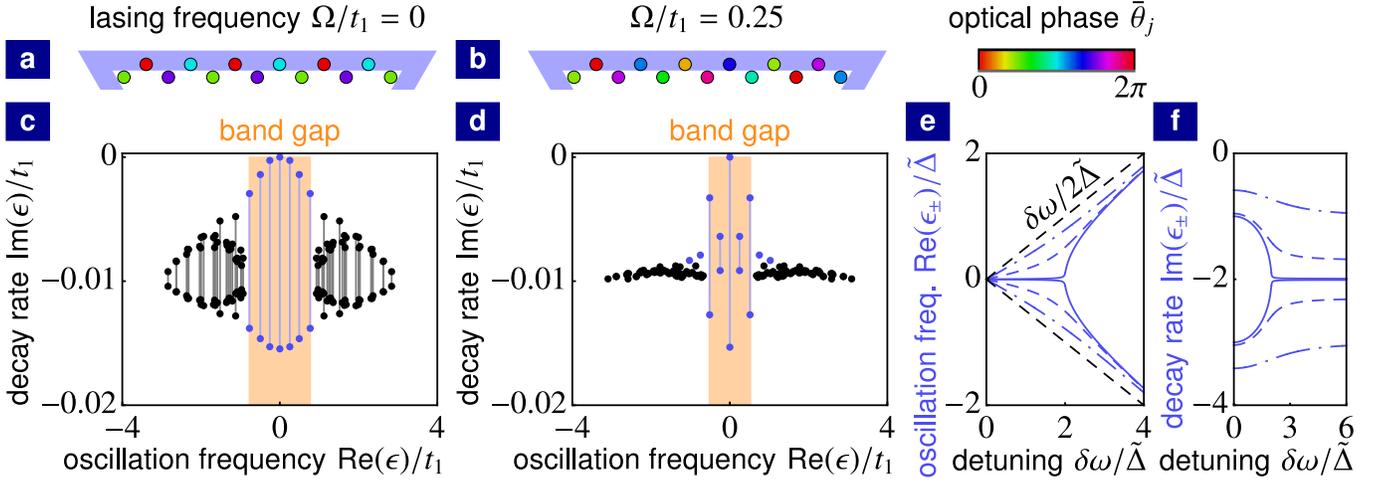}
\caption{Elementary excitations for lasing of different edge modes. (a) and (b) Mean-field steady-state distribution of optical phases along the top edge of the photonic array for the lasing frequency $\Omega/t_1 = 0$ and $\Omega/t_1=0.25$, respectively. (c) and (d) Complex spectrum of elementary excitations with band gap (orange region), bulk modes (black points) as well as edge modes (blue points) for the lasing frequency $\Omega/t_1 = 0$ and $\Omega/t_1=0.25$, respectively. Gray and blue lines show the splitting in imaginary parts of complex frequencies due to the hybridization of bulk modes and edge modes, respectively. (e) and (f) Real part and imaginary part, respectively, of complex frequencies $\epsilon_{\pm}$ for two hybridized modes as a function of the detuning $\delta\omega$ for different values of the decay-rate difference $\delta \Gamma/\tilde{\Delta} = 0.02$ (full lines), $\delta \Gamma/\tilde{\Delta} = 0.6$ (dashed lines) and $\delta \Gamma/\tilde{\Delta} = 2$ (dot-dashed lines). (Parameters: (a-d) $t_2/t_1=0.15$, $\phi=\pi/2$, $\gamma//t_1=0.01$, $g/t_1=0.05$, $N=61$; (e) and (f) $ \bar{\omega}/\tilde{\Delta} = 0$, $ \bar{\Gamma}/\tilde{\Delta} = -2$)}
\label{excitations}
\end{figure*}

We now discuss the coupling of modes, $\mathcal{E}_p^{(m)}$ and $\tilde{\mathcal{E}}_p^{(n)}$, from the two different branches due to the off-diagonal blocks $\tilde{\mathbf{\Delta}} $ and $\tilde{\mathbf{\Delta}}^*$ of the dynamical matrix. The coupling between modes $\mathcal{E}_p^{(m)}$ and $\tilde{\mathcal{E}}_p^{(n)}$ is described by the $2\times2$ dynamical matrix
\begin{equation}\label{dyn 2}
\mathcal{\tilde{D}}^{(m,n)}  = \begin{pmatrix}
\omega_m - \Omega & 0 \\
0& -\omega_n  + \Omega 
\end{pmatrix}
+
i \begin{pmatrix}
\tilde{\Gamma}_{mm}  & \tilde{\Delta}_{mn} \\
\tilde{\Delta}_{nm}^* & \tilde{\Gamma}_{nn}
\end{pmatrix},
\end{equation}
if their frequencies are isolated from the rest of the passive-system spectrum, i.e.~$|\omega_{m/n}+\omega_q-2\Omega|\gg|\tilde{\Delta}_{m/nq}|$, $|\omega_{m/n}-\omega_q|\gg\tilde{\Gamma}_{m/nq}$ for all $q\neq m,n$.
The frequencies of the edge modes in the band gap of the passive system satisfy this condition for moderate system sizes and for $t_1,t_2\gg g,\gamma$ as considered in this manuscript. Diagonalizing the $2\times2$ dynamical matrix, we obtain complex frequencies of hybridized modes
\begin{equation}\label{hyb e}
\epsilon_{\pm}^{(m,n)} = \bar{\omega}_{mn} - i \bar{\Gamma}_{mn} \pm \frac{1}{2}\sqrt{\left(\delta\omega_{mn} + i \delta\Gamma_{mn} \right)^2 - 4|\tilde{\Delta}_{mn} |^2},
\end{equation}
where $\bar{\omega}_{mn} = \left(\omega_m - \omega_n\right)/2$, $\delta\omega_{mn} =  \omega_m + \omega_n -2\Omega$, $\bar{\Gamma}_{mn} = - \left(\tilde{\Gamma}_{mm} + \tilde{\Gamma}_{nn}\right)/2$, and
$\delta\Gamma_{mn} =  \tilde{\Gamma}_{mm}-\tilde{\Gamma}_{nn}$. 
The real parts of complex frequencies correspond to oscillation frequencies and the imaginary parts of complex frequencies correspond to decay rates or amplification rates. The real part and the imaginary part of the complex frequencies $\epsilon^{(m,n)}_{\pm}$ are shown in Figs.~\ref{excitations}e and \ref{excitations}f, respectively, as a function of the detuning $\delta\omega_{mn}$. One can see that due to the anti-Hermitian coupling of passive-system normal modes, the real parts of the complex frequencies are attracted to each other, $\textrm{Re}\left(\epsilon^{(m,n)}_{+} - \epsilon^{(m,n)}_{-}\right)< |\delta\omega_{mn}|$. On the other hand, the imaginary parts of the complex frequencies split. This is an example of level attraction, which is a general concept appearing in various physical platforms \cite{savvidis2001,bernier2014,bernier2018}. For $2\bar{\Gamma}_{mn} > \sqrt{\delta\Gamma_{mn}+4|\tilde{\Delta}|^2}$, both hybridized modes decay as the imaginary parts of the complex frequencies are negative. The splitting in the imaginary parts of the complex frequencies is large for small detunings $\delta\omega$ leading to a slowly-decaying mode and a fast-decaying mode. For a large detuning $|\delta\omega_{mn}| \gg |\tilde{\Delta}_{mn}|$, the hybridization is negligible and the frequencies of uncoupled modes $\epsilon_{+}^{(m,n)} \approx \omega_m -\Omega + i \tilde{\Gamma}_{mm}$ as well as $\epsilon_{-}^{(m,n)} \approx - \omega_n +\Omega + i \tilde{\Gamma}_{nn}$ are recovered.

The hybridization of two edge modes from the two different branches described by the $2\times2$ dynamical matrix $\mathcal{\tilde{D}}^{(m,n)} $ will be shown in the next section to have important consequences for the complex spectrum of elementary excitations.

\section{Spectrum of elementary excitations}\label{sec spectrum}

We now investigate the complex spectrum of elementary excitations in the regime $t_1,t_2 \gg g,\gamma$. In Fig.~\ref{excitations}d, we plot the complex spectrum of elementary excitations for lasing of the edge mode with the frequency $\Omega/t_1 = 0.25$. This spectrum reveals generic features of elementary excitations in topological lasers. 

Normal modes of elementary excitations are formed from either bulk modes (black points) or edge modes (blue points) of the passive system. In the regime $t_1,t_2 \gg g,\gamma$, the oscillation frequencies (real parts of complex frequencies) of elementary excitations are predominantly determined by the eigenfrequencies of the Hermitian part $\mathcal{H}$, which consists of two branches $\epsilon_p^{(m)} = \omega_m - \Omega$ and $ \tilde{\epsilon}_p^{(m)} = - \omega_m + \Omega$ formed from the passive system frequencies $\omega_m$. These two branches are shifted in respect to each other by the lasing frequency $\Omega$. Since the lasing frequency lies in the passive-system band gap, the band gaps of the two branches overlap, giving rise to a band gap in the spectrum of elementary excitations (orange region in Fig.~\ref{excitations}d). The band gap in the spectrum of elementary excitations represents a range of frequencies, within which no bulk modes are excited by elementary excitations. As the lasing frequency $\Omega/t_1=0.25$ does not lie in the middle of the passive-system band gap, the band gaps of the two branches overlap only partially. As a result, the band gap in the spectrum of elementary excitations is smaller than that of the passive system.

All imaginary parts of complex frequencies are negative (except from a single frequency with a vanishing imaginary part discussed later) confirming the stability of the steady state. For moderate system sizes that we consider here, $|\omega_m - \omega_n|\gg g,\gamma$ for all $m\neq n$ and edge-mode frequencies $\omega_m$ lying in the band gap of the passive system. As a result, every edge mode $\mathcal{E}_p^{(m)}$ can significantly hybridize only with a single mode $\tilde{\mathcal{E}}_p^{(n)}$ from the other branch and their coupling is described by the $2\times2$ dynamical matrix (\ref{dyn 2}). Due to the large spatial overlap of edge modes in the pumped region $\mathbb{P}_j$, the coupling $|\tilde{\Delta}_{mn}|$ between edge modes overcomes the detuning of their passive-system frequencies $|\delta\omega_{mn}|$. This leads to a large hybridization of edge modes and to a distinctive splitting in the imaginary parts of their complex frequencies (blue lines  in Fig.~\ref{excitations}d).

Two passive-system normal modes formed from the lasing mode $\mathbf{e}^{(l)}$ are always degenerate at frequency $\epsilon_p^{(l)}=\tilde{\epsilon}_p^{(l)} = 0$. The hybridization of these two modes gives rise to a non-decaying mode with the complex frequency $\epsilon^{(l,l)}_{+} = 0 $ and a fast-decaying mode with the complex frequency $\epsilon^{(l,l)}_{-} = -2i\bar{\Gamma}_{ll}$ (see Appendix~B for more details). These non-decaying and fast-decaying excitations correspond to undamped fluctuations in the phase of the lasing mode and largely-damped fluctuations in the amplitude of the lasing mode, respectively, which are characteristic for a laser driven above threshold \cite{gardiner2004}.

Note that the hybridization of edge modes $\mathcal{E}_p^{(m)}$ and $\mathcal{E}_p^{(n)}$ from the same branch is negligible because the detuning of passive-system frequencies $|\delta\omega_{mn}|=|\omega_m - \omega_n|$ is always larger than the coupling term $|\tilde{\Gamma}_{mn}|$ between these modes.

Since couplings $\tilde{\Delta}_{nm}$ and $\tilde{\Gamma}_{nm}$ between bulk modes are small, the hybridization of bulk modes is typically also negligible. Complex frequencies of non-hybridized bulk modes (black points in Fig.~\ref{excitations}d) acquire imaginary parts  $\textrm{Im}\, \epsilon^{(m)} \approx -\gamma$ and $\textrm{Im} \,\tilde{\epsilon}^{(m)} \approx -\gamma$ due to the diagonal term  $\tilde{\Gamma}_{mm}\approx-\gamma$ in the anti-Hermitian part of the dynamical matrix.

Note that for the value of the Haldane flux $\phi=\pi/2$, a small hybridization of bulk modes occurs for lasing at the frequency $\Omega/t_1=0$ (see Fig.~\ref{excitations}c), because bulk modes are pairwise degenerate due to the symmetry, $\mathbf{S} \,\mathbf{H}\,\mathbf{S} = - \mathbf{H}^*$, of the passive-system Hamiltonian $\mathbf{H}$, where $\mathbf{S}$ is a unitary and  $\mathbf{S}^2=\mathbb{1}$ (see Appendix~C for more details). However, the splitting in imaginary parts of complex frequencies for bulk modes is small in comparison to the splitting for edge modes and the hybridization of bulk modes does not appear for other values of the Haldane flux $\phi\neq\pi/2$  or for other lasing frequencies $\Omega/t_1\neq0$.

\section{Long-lived elementary excitations}\label{llee}

We now discuss long-lived elementary excitations, which occur in the Haldane model for lasing at a frequency lying in the middle of the passive-system band gap (vicinity of $\Omega/t_1=0$ for $\phi\approx\pi/2$).

In Fig.~\ref{excitations}c, we plot the complex spectrum of elementary excitations for lasing at the frequency $\Omega/t_1=0$, which lies in the middle of the passive-system band gap. Long-lived elementary excitations with decay rates, which are orders of magnitude smaller than any other energy scale in the system ($\gamma$, $g$, $t_1$ and $t_2$), appear due to a large hybridization of edge modes. 
The very slow decay of long-lived elementary excitations leads to an ultra slow relaxation of the topological laser towards the mean-field steady state, which was numerically observed in Ref.~\cite{secli2019}. In contrast to lasing at the frequency $\Omega/t_1=0$, decay rates of slowly-decaying modes are comparable to $\gamma$ for the lasing at the frequency $\Omega/t_1=0.25$ (see Fig.~\ref{excitations}d).

To understand the dependence of the spectrum for elementary excitations on the selection of a lasing edge mode, we can expand the edge-mode frequencies $\omega_m = \Omega + v_1 \left( m - l \right) + v_2 \left(m-l\right)^2 + \mathcal{O}\left(\left(m-l\right)^3\right)$ around the lasing frequency $\Omega$, where the index $l$ labels the lasing mode. For $|v_1|\gg |v_2|$, the frequency of edge mode $m$ is close to the frequency of edge mode $2l-m$ from the other branch of passive-system normal modes and their detuning is $\delta \omega_{m(2l-m)} = 2 v_2\left(m-l\right)^2 + \mathcal{O}\left(\left(m-l\right)^4\right)$. If the nonlinear coefficient $|v_2|$ and, as a consequence, also the detuning $\delta \omega_{m(2l-m)}$ are small compared to the coupling $|\tilde{\Delta}_{m(2l-m)}|$ between the edge modes, the edge modes significantly hybridize giving rise to a large splitting in the imaginary part of the complex frequencies, see Eq.~(\ref{hyb e}) and Fig.~\ref{excitations}f. On the other hand, if the nonlinear coefficient  $|v_2|$ is comparable to or larger than the coupling $|\tilde{\Delta}_{m(2l-m)}|$, the resulting detuning $\delta \omega_{m(2l-m)}$ obstructs the hybridization and the splitting in the imaginary parts of edge-mode frequencies is reduced.

\begin{figure*}[htbp]
\centering
\includegraphics[width=\linewidth]{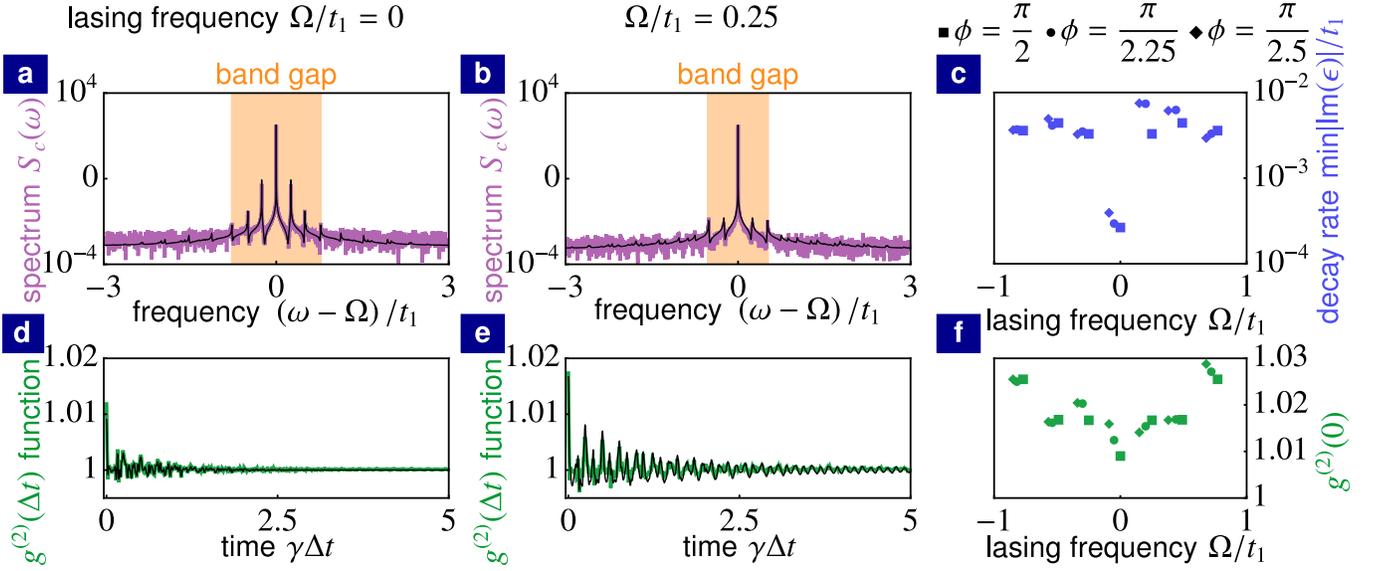}
\caption{Coherence properties of topological laser. (a) and (b) Optical spectrum of a pumped optical site lying at the edge of the topological array for lasing at the frequency $\Omega/t_1 =0 $ and at the frequency $\Omega/t_1 =0.25 $, respectively. Linearization of Langevin equations around the mean-field steady state (black line) and numerical simulations of nonlinear Langevin equations (purple line). The orange region shows the band gap in the spectrum of elementary excitations. (d) and (e) Second-order autocorrelation function of a pumped optical site lying at the edge of the topological array for lasing at the frequency $\Omega/t_1 =0 $ and at the frequency $\Omega/t_1 =0.25 $, respectively.  Linearization of Langevin equations around the mean-field steady state (black line) and numerical simulations of nonlinear Langevin equations (green line). (c) Smallest decay rate of elementary excitations $\textrm{min}_{\alpha\neq\eta}|\textrm{Im}\,\epsilon^{(\alpha)}|$ as a function of the lasing frequency $\Omega$ for Haldane flux $\phi=\pi/2$ (squares), $\phi=\pi/2.25$ (circles) and $\phi = \pi/2.5$ (diamonds). (f) Equal-time second-order autocorrelation function $g^{(2)}(0)$ as a function of the lasing frequency $\Omega$ for Haldane flux $\phi=\frac{\pi}{2}$ (squares), $\phi=\frac{\pi}{2.25}$ (circles) and $\phi = \frac{\pi}{2.5}$ (diamonds). (Parameters: $t_2/t_1=0.15$, $\gamma/t_1=0.01$, $g/t_1=0.05$, $I_{\rm sat}\gamma/q=25$; (a), (b), (d) and (e) $\phi=\pi/2$)
}
\label{coherence}
\end{figure*}

For the Haldane model, the dispersion of edge-mode frequencies is linear in the middle of the passive-system band gap for any $\phi\neq0,\pi$. As $v_2$ is very small for lasing at a frequency lying in the middle of the passive-system band gap, long-lived elementary excitations, whose decay rate is orders of magnitude smaller than any other energy scale in the system ($\gamma$, $g$, $t_1$ and $t_2$), appear for any value of the Haldane flux. This can be seen in Fig.~\ref{coherence}c, where we plot the smallest decay rate $\textrm{min}_{\alpha\neq\eta}|\textrm{Im}\,\epsilon^{(\alpha)}|$ (index $\eta$ labels the non-decaying mode) as a function of the lasing frequency $\Omega$ for different values of the Haldane flux. On the other hand, for lasing at any frequency, which does not lie in the middle of the passive-system band gap, the nonlinear coefficient $v_2$ is large enough to give rise to a considerable detuning of edge mode frequencies compared to the coupling of edge modes. The hybridization of edge modes is then obstructed and   the smallest decay rate is comparable to $\gamma$, see  Fig.~\ref{coherence}c.

Long-lived elementary excitations are not unique to topological lasers.
They generically appear in one-dimensional arrays if the dispersion of the passive frequencies is linear around the lasing frequency, see Appendix~D for more details. However, in contrast to topological lasers, long-lived elementary excitations in topologically trivial lasers are sensitive to disorder. Even moderate disorder in the on-site frequencies $\nu_j$ can obstruct or enhance the hybridization of the passive-system normal modes, leading to a large change in the decay rate of the long-lived elementary excitations as well as in their oscillation frequency, see Appendix~G.
On the other hand, long-lived elementary excitations in a topological laser ($\phi\neq0,\pi$) are robust against moderate on-site disorder. Their decay rate is only marginally affected and thus it remains orders of magnitude smaller than $\gamma$ in the presence of disorder (see Appendix~F for more details). The oscillation frequency of long-lived elementary excitations is unaffected by the disorder. The long-lived elementary excitations are robust against disorder since they appear due to the hybridization of topological edge modes, which are protected against disorder as long as disorder is not strong enough to close the topological band gap \cite{hafezi2013,harari2018}.

In general, long-lived elementary excitations appear due to the hybridization of edge modes for any topological model with a linear dispersion of edge-mode frequencies. The lifetime of elementary excitations can be suppressed by selecting a lasing frequency around which the dispersion of edge-mode frequencies is no longer linear. The long-lived elementary excitations will be shown in the next section to have crucial consequences for light coherence.

\section{Coherence properties}

We now discuss coherence properties of topological lasers and how they are influenced by long-lived elementary excitations, investigating the emission spectrum of pumped optical sites and the second-order autocorrelation function. 

We start by studying the autocorrelation of complex optical amplitudes $\langle c_j(t) c_j^*(t+\Delta t)\rangle$. The dominant contribution in this autocorrelation is determined by phase fluctuations $\delta \theta_j$, where $c_j = \left(\bar{C}_j + \delta C \right)\,e^{i\left(-\Omega t + \bar{\theta}_j + \delta\theta_j\right)}$ and $\bar{c}_j = \bar{C}_j e^{i\bar{\theta}_j}$ (see Appendix~E for more details). The amplitude fluctuations $\delta C_j$ are negligible in comparison to the large mean-field occupation $\bar{C}_j^{2}$ \cite{gardiner2004}. Amplitude fluctuations $\delta C_j$ and phase fluctuations $\delta \theta_j$ are linearly related to the fluctuations of complex amplitudes $ \delta c_j $ and $ \delta c_j^* $ as well as to the normal modes of elementary excitations
\begin{equation}
\begin{pmatrix}
\delta\mathbf{ C} \\
\delta \mathbf{\Theta}
\end{pmatrix}
=
\mathcal{W} \mathcal{N},
\end{equation}
where $\delta\Theta_j = \bar{C}_j\,\delta\theta_j $, the vector $\mathcal{N}$ contains the complex amplitudes of the normal modes and $\mathcal{W}$ is the transformation matrix. This allows us to express the autocorrelations of complex optical amplitudes
\begin{gather}
 \langle c_j(t) c_j^*(t + \Delta t)\rangle \approx \bar{C}_j^{2}\,e^{i\Omega\Delta t - |\Delta t|/\tau_c} \nonumber \\
 + \sum_{\alpha\neq\eta}n_{\alpha}|\mathcal{W}_{(j+N)\alpha}|^2\,e^{i\left(\textrm{Re}\,\epsilon^{(\alpha)}+\Omega\right)\Delta t + \textrm{Im}\,\epsilon^{(\alpha)}|\Delta t|}\label{spec ee}
\end{gather}
in terms of the complex frequencies of elementary excitations $\epsilon^{(\alpha)}$ and the occupations $n_{\alpha} = \frac{1}{2|\textrm{Im}\,\epsilon^{(\alpha)}|}\left(\mathcal{R}\mathcal{R}^{\dagger}\right)_{\alpha\alpha}$ of the corresponding normal modes, where $\tau_c = 2\bar{C}_j^2|\mathcal{W}_{(j+N)\eta}|^{-2}/\left(\mathcal{R}\mathcal{R}^{\dagger}\right)_{\eta\eta}$ is a coherence time, the index $\eta$ labels the non-decaying mode, and $\mathcal{R}\mathcal{R}^{\dagger}=\frac{1}{2} \mathcal{W}^{-1}\mathcal{Q}\mathcal{Q}^{\dagger}\left(\mathcal{W}^{-1}\right)^{\dagger}$ is the correlation matrix for the normal modes (see Appendix~E for a detailed derivation).

 The optical spectrum $S_{c_j}(\omega)$ is the Fourier transform of the autocorrelation $\langle c_j(t) c_j^*(t+\Delta t)\rangle$. The light field emitted by optical sites is proportional to the complex amplitudes of the optical sites as described by input-output formalism \cite{gardiner2004}. As a result, the emission spectrum is proportional to the optical spectrum $S_{c_j}(\omega)$. The optical spectrum of a pumped optical site located at the edge of the array is shown in Figs.~\ref{coherence}a and \ref{coherence}b for lasing at the frequency $\Omega/t_1=0$ and $\Omega/t_1=0.25$, respectively. We find a good quantitative agreement between the optical spectrum determined from the linearized Langevin equations (black lines), see Eq.~(\ref{spec ee}), and the optical spectrum obtained from numerical simulations (purple line) of the nonlinear Langevin equations (\ref{full eq}). The ensemble average $\langle c_j(t) c_j^*(t + \Delta t)\rangle$ can be replaced by a time average for a steady-state laser operation. 

For lasing at both frequencies, the optical spectrum contains a central peak at the lasing frequency corresponding to the first term in Eq.~(\ref{spec ee}). Undamped fluctuations in the phase of the lasing mode associated with the non-decaying normal mode of elementary excitations lead to a phase diffusion of light field, giving rise to a Lorentzian shape of the central peak with a linewidth $2/\tau_c$ \cite{gardiner2004}.  The linewidth is proportional to the strength of fluctuations $q$ as well as inversely proportional to the number of pumped sites and the occupation of the pumped optical site $\bar{C}_j^2$. The linewidth is approximately constant for lasing of any edge mode. Small deviations in the linewidth occur due to moderate discrepancies in the spatial profile $|\mathcal{W}_{j\alpha}|^2$ of individual edge modes.

For lasing at the frequency $\Omega/t_1=0$ lying in the middle of the passive-system band gap, the optical spectrum contains also satellite peaks (see Fig.~\ref{coherence}a). The satellite peaks appear due to the incoherent population of normal modes for elementary excitations, corresponding to the terms on the second line of Eq.~(\ref{spec ee}). The occupation $n_{\alpha}$ of normal modes for elementary excitations is inversely proportional to the decay rate $|\textrm{Im}\,\epsilon^{(\alpha)}|$ and proportional to the strength of noise $q$.
As a result, long-lived elementary excitations with a very small decay rate are largely populated giving rise to the satellite peaks in the optical spectrum. This large incoherent population of normal modes for elementary excitations leads to large phase fluctuations in the emitted light field decreasing its coherence.

 Since elementary excitations are not dependent on the absolute scaling, $I_{\rm sat}$, of the mean-field steady-state solution, the occupation of normal modes $n_{\alpha}$ does not depend on the mean number of photons in the lasing mode $\bar{n}$. As a result, the height of the satellite peaks is also independent of the the mean number of photons in the lasing mode $\bar{n}$. 
 
 Our results show that large phase fluctuations and the decreased light coherence of emitted light field persist even when moderate on-site disorder is introduced (see Appendix~F). This is due to the robustness of edge modes and their frequencies against disorder. As a result, long-lived elementary excitations with a very small decay rate and a large occupation $n_{\alpha}$ of the corresponding normal modes occur even if moderate on-site disorder is considered.

On the other hand, the incoherent population of elementary excitations and corresponding phase fluctuations can be suppressed by selecting a different lasing frequency. As it was shown in the previous section, the lifetime of elementary excitations is reduced by at least one order of magnitude for lasing at a frequency, which does not lie in the middle of the passive-system band gap, see Fig.~\ref{coherence}c. As a result, the incoherent population of elementary excitations and  the corresponding satellite peaks in the optical spectrum are suppressed (see Fig.~\ref{coherence}b) leading to a larger coherence of emitted light than for $\Omega/t_1 = 0$. 

The second-order autocorrelation function $g^{(2)}_j$ describes correlations in the intensity of emitted light at different times \cite{gardiner2004}. For a laser, it is desired that these intensity correlations vanish corresponding to $g^{(2)}_j=1$. The second-order autocorrelation function is determined by amplitude fluctuations \cite{gardiner2004}
\begin{gather}
g^{(2)}_j (\Delta t)= \frac{\langle c_j(t) c_j(t+\Delta t)c^*_j(t+\Delta t)c^*_j(t) \rangle}{\langle c_j(t) c^*_j(t)\rangle\langle c_j(t + \Delta t)c^*_j(t + \Delta t)\rangle}\nonumber\\
= 1 + \frac{4}{\bar{C}^2_j}\langle \delta C_j (t) \delta C_j(t + \Delta t) \rangle + \mathcal{O}\left(\frac{1}{\bar{C}^{4}_j}\right). \label{g2 eq}
\end{gather}
Amplitude autocorrelations $\langle \delta C_j (t) \delta C_j(t + \Delta t) \rangle$ can be expressed in terms of normal modes’ correlations (see Appendix~E).
The second-order autocorrelation function $g^{(2)}_j (\Delta t)$ for a pumped optical site located at the edge of the array is shown in Figs.~\ref{coherence}d and \ref{coherence}e for lasing at the frequency $\Omega/t_1=0$ and $\Omega/t_1=0.25$, respectively. We compare the results determined from the linearized Langevin equations (black line) to numerical simulations (green line) of the nonlinear Langevin equations (\ref{full eq}). The ensemble average $\langle c_j(t) c_j(t + \Delta t)c^*_j(t + \Delta t)c^*_j(t) \rangle$ can be replaced by a time average for a steady-state laser operation. 

For lasing at both frequencies, the equal-time second-order autocorrelation function  $g^{(2)}_j (0)$ is close to unity as expected for a laser, which is driven well above threshold. With the time difference $\Delta t$, $g^{(2)}_j (\Delta t)$ decays to unity at time comparable to $1/\gamma$. This shows that amplitude fluctuations correspond to fast-decaying elementary excitations. We can see that $g^{(2)}_j (0)$ and temporal oscillations of $g^{(2)}_j (\Delta t)$ are larger for lasing at the frequency $\Omega/t_1 = 0.25$ than for lasing at the frequency $\Omega/t_1 = 0$ lying in the middle of the band gap. 

 In Fig.~\ref{coherence}f, we plot $g^{(2)}_j (0)$ as a function of the lasing frequency $\Omega$ for different values of the Haldane flux. One can see that $g^{(2)}_j (0)$ is, in general, moderately larger for lasing at a frequency which does not lie in the middle of the band gap for all values of the Haldane flux. This shows that lasing at these frequencies leads to moderately larger amplitude fluctuations.

\section{Experimental parameters}
We estimate parameters of our model (\ref{full eq}) to be relevant for recent experiments \cite{bandres2018}. Typical parameters for arrays of coupled microring resonators are the decay rate $\gamma\sim1\,\textrm{GHz}$ and the hopping amplitude $t_1\sim100\,\textrm{GHz}$ with a feasible ratio $\gamma/t_1\sim0.01$ \cite{hafezi2013}. Based on the Haldane model (see Fig.~\ref{steady}a) with the group velocity of edge modes $v_g/t_1 \sim 1$ (in units of the lattice constant), we can estimate that the frequency separation  $|\omega_m-\omega_n|\sim\gamma,g$ of edge modes  $m\neq n$ is comparable to their linewidth  $\gamma$ and gain $g$ for a total number of microring resonators $N\sim10^4$. For $N\sim100$  (as implemented in Ref.~\cite{bandres2018}), $|\omega_m-\omega_n|\gg g,\gamma$.
As a result, each edge mode can distinctively hybridize only with one edge mode from the other branch of passive-system frequencies as the anti-Hermitian coupling to all other modes is negligible compared to their large frequency separation.

The dominant source of noise 
is the spontaneous emission at rate $q\sim 100\,{\rm GHz}$ \cite{hodaei2014}. A typical circulating power in the lasing mode of a single microring resonator is  $P_c \sim 1 \,{\rm mW}$ which corresponds to a typical number of photons $\bar{n} \sim 10^3$ in the lasing mode \cite{hodaei2014}. We conclude that our model with $I_{\rm sat}\gamma/q\sim10$ describes an experimentally-relevant relative strength of noise compared to the number of photons in the lasing mode.

\section{Conclusions}
We have demonstrated that long-lived elementary excitations, which emerge due to the hybridization of topological edge modes, lead to large phase fluctuations and a decrease in the coherence of the emitted light field. In contrast to long-lived elementary excitations in a trivial laser, the decay rate and the oscillation frequency of long-lived elementary excitations in a topological laser are robust against disorder. Even though we focus in our manuscript on the Haldane model, long-lived elementary excitations appear for any topological model if the dispersion of edge-mode frequencies is approximately linear around the lasing frequency.
Our results for the Haldane model show that the deviation from a linear dispersion around lasing frequencies, which do not lie in the middle of the passive-system band gap, is sufficient to obstruct the hybridization of edge modes. As a result, the lifetime of elementary excitations is reduced by orders of magnitude and the phase fluctuations are largely suppressed. On the other hand, this leads to a moderate increase of amplitude fluctuations and the second-order autocorrelation function. However, the second-order autocorrelation function still remains close to unity. In the future, different topological models can be studied to provide insight into how elementary excitations in topological lasers are affected by the presence of several topological band gaps supporting edge modes with opposite chirality \cite{ozawa2019}, a pseudospin degree of freedom in pseudo quantum spin Hall systems \cite{hafezi2011,hafezi2013} or topological lasing in synthetic dimensions \cite{lustig2019}. 

\appendix

\section*{Appendix A: Linearization of Langevin euqations}
\label{app gold}
\setcounter{equation}{0}
\renewcommand{\theequation}{A{\arabic{equation}}}

We now derive the linear Langevin equations (\ref{bog eq}) for elementary excitations around the mean-field steady state. To this end we substitute the decomposition of the complex optical amplitude $c_j = \left(\bar{c}_j + \delta c_j \right)e^{-i\Omega t}$ into the full nonlinear Langevin equations (\ref{full eq}). Omitting second- and higher-order terms in optical modulations $\delta c_j$, we obtain the linearized Langevin equations
\begin{align}
i \frac{\rm d}{{\rm d} t}\delta c_j =& -\Omega\delta c_j  + \sum^{N}_{k=1} H_{jk} \delta c_k  + i \Gamma_{jj}\delta c_j + i\Delta_{jj}\delta c^*_j\nonumber\\
 &+ Q_{jj} c_{j,{\rm in}}e^{i\Omega t},\label{eq lin 1}\\
i \frac{\rm d}{{\rm d} t}\delta c^*_j =&\,\Omega\delta c^*_j - \sum^{N}_{k=1} H^*_{jk} \delta c^*_k + i\Delta_{jj}\delta c_j +i \Gamma_{jj} \delta c^*_j\nonumber\\
& - Q_{jj} c^*_{j,{\rm in}}e^{-i\Omega t},\label{eq lin 2}
\end{align}
where
\begin{equation}
\Gamma_{jj} =  -\gamma + \frac{\mathbb{P}_{j}\,g}{\left(1 + \frac{|\bar{c}_{j}|^2}{I_{\rm sat}}\right)^2},\,\,\,\,\Delta_{jj} =  -  \frac{\mathbb{P}_{j}\,g\,\frac{\bar{c}_{j}^2}{I_{\rm sat}}}{\left(1 + \frac{ |\bar{c}_{j}|^2}{I_{\rm sat}}\right)^2},
\end{equation}
and we used that
\begin{equation}
\left[ - i \gamma + i \frac{\mathbb{P}_j g}{1 + \frac{|\bar{c}_j|^2}{I_{\rm sat}}} \right]\bar{c}_j  + \sum^{N}_{k=1} H_{jk} \bar{c}_k=0.
\end{equation}
(\ref{eq lin 1}) and (\ref{eq lin 2}) can be written in form of a matrix equation (\ref{bog eq}).

\section*{Appendix B: Non-decaying mode}
\label{app gold}
\setcounter{equation}{0}
\renewcommand{\theequation}{B{\arabic{equation}}}

Here we discuss the hybridization of two passive-system normal modes which are formed from the lasing mode, giving rise to the non-decaying mode. We label the lasing mode by the index $l$. Since $\omega_l = \Omega$, the pair of passive-system normal modes is degenerate $\epsilon_p^{(l)}=\tilde{\epsilon}_p^{(l)} = 0$ leading to a large hybridization of the pair. The lasing mode coincides with the mean-field steady-state solution $e_j^{(l)} \approx \bar{c}_j\, e^{i\varphi}/\sqrt{\bar{n}}$, 
where $\bar{n}$ is the mean number of photons in the lasing mode and $\varphi$ is an arbitrary phase. This gives 
\begin{equation}
\bar{\Gamma}_{ll} = \tilde{\Delta}_{ll}\,e^{2i\varphi} = g\sum_{j=1}^N \mathbb{P}_{j}\,\frac{\frac{|\bar{c}_{j}|^4}{I_{\rm sat}\bar{n}}}{\left(1 + \frac{|\bar{c}_{j}|^2}{I_{\rm sat}}\right)^2},
\end{equation}
and $\delta\Gamma_{ll} $ trivially vanishes. Using also $\bar{\omega}_{ll}=0$ and $\delta\omega_{ll}=0$, we see from Eq.~(\ref{hyb e}) that the hybridization of this mode pair gives rise a non-decaying mode with the complex frequency $\epsilon^{(l,l)}_{+} = 0 $ and a fast-decaying mode with the complex frequency $\epsilon^{(l,l)}_{-} = -2i\bar{\Gamma}_{ll}$. 

\section*{Appendix C: Haldane flux $\phi=\pi/2$}
\label{app phi0}
\setcounter{equation}{0}
\renewcommand{\theequation}{C{\arabic{equation}}}
The value of the Haldane flux $\phi=\pi/2$ represents a special case because the Hamiltonian $\mathbf{H}$ of the passive system then exhibits the following symmetry $\mathbf{S} \,\mathbf{H}\,\mathbf{S} = - \mathbf{H}^*$, where $\mathbf{S}$ is a unitary matrix and  $\mathbf{S}^2=\mathbb{1}$. The unitary transformation $\mathbf{S}$ introduces the phase shift $\pi$ between the two sublattices of the Haldane model, i.e.~$c_j\rightarrow c_j$ for sublattice $A$ and $c_j\rightarrow-c_j$ for sublattice $B$. Due to this symmetry, the spectrum of the passive system consists of a zero frequency and frequency pairs $(\omega_m, \omega_{\tilde{m}})$, $\omega_{\tilde{m}} = - \omega_{m} $. As a result, all passive-system normal modes are pairwise degenerate for $\Omega/t_1=0$ leading to the hybridization of all degenerate pairs described by the $2\times2$ dynamical matrix $\tilde{\mathcal{D}}^{(m,\tilde{m})}$, see Eq.~(\ref{dyn 2}). As the unitary $\mathbf{S}$ introduces only a local phase shift, $|e^{(m)}_j| = |e^{(\tilde{m})}_j| $ leading to $\bar{\Gamma}_{m\tilde{m}}=\tilde{\Gamma}_{mm}=\tilde{\Gamma}_{\tilde{m}\tilde{m}}$ and $\delta\Gamma_{m\tilde{m}}=0$. Using also $\delta\omega_{m\tilde{m}}=0$,  the complex frequencies of hybridized modes are $\epsilon^{(m,\tilde{m})}_{\pm} = \omega_m + i\left(\bar{\Gamma}_{mm}\pm|\tilde{\Delta}_{m\tilde{m}}|\right)$. Since the coupling $\tilde{\Delta}_{mn}$ between bulk modes is small, the hybridization leads to a small splitting in imaginary parts of complex frequencies for bulk modes (gray lines in Fig.~\ref{excitations}c). On the other hand, the large coupling of edge modes leads to a large splitting in the imaginary parts of complex frequencies (blue lines in Fig.~\ref{excitations}c).

\section*{Appendix D: One-dimensional laser array}
\label{disorder}
\setcounter{equation}{0}
\renewcommand{\theequation}{D{\arabic{equation}}}
\renewcommand{\thefigure}{D{\arabic{figure}}}

In this appendix, we discuss how long-lived elementary excitations generically appear in one-dimensional laser arrays. We consider a one-dimensional array with $N$ optical sites, whose complex amplitudes are described by the Langevin equations (\ref{full eq}). All optical sites in the one-dimensional array are pumped,~i.e. $\mathbb{P}_j=1$ for all $j$. We do not consider any particular Hamiltonian $H_{jk}$. We only assume periodic boundary conditions $c_{N+j} = c_j$, that  $\nu_j = \nu = {\rm const}$, and that the Hamiltonian $H_{jk}$ is translationally invariant,~i.e. that the Hamiltonian is invariant under the transformation $c_j\rightarrow c_{j+R}$ for any integer $R$. In this case, the passive-system normal modes are plane waves with complex amplitudes $\mathcal{N}_m = \frac{1}{\sqrt{N}}\sum_{j=1}^{N}e^{ijm}c_j$. The index $m$ represents a quasi-momentum and it spans values $m=\frac{2\pi}{N},\frac{4\pi}{N},...,2\pi$. The mean-field equations of motion (omitting stochastic terms in the Langevin equations) for the complex amplitudes $\mathcal{N}_m $ are 
\begin{align}
i\frac{\rm d}{{\rm d}t} \mathcal{N}_m =& \left(\omega_m - i\gamma\right)\mathcal{N}_m\nonumber\\ 
&+ \sum_{j=1}^N\,\frac{ i \frac{g}{N}\sum_n e^{ij(m-n)}\mathcal{N}_n}{1+\sum_{n,o}e^{-ij(n-o)}\mathcal{N}_n\mathcal{N}^*_o/NI_{\rm sat}},
\end{align}
where $\omega_m$ is the oscillation frequency of the normal mode $\mathcal{N}_m$. These mean-field equations of motion have a stationary solution $\bar{\mathcal{N}}_m = \delta_{ml}\sqrt{NI_{\rm sat}\left(\frac{g}{\gamma}-1\right)}\,e^{-i\varphi}$ corresponding to the lasing of a single mode $l$, where $\varphi$ is an arbitrary phase. Note that a stationary solution exists for any normal mode $l$ lasing. Switching back to the basis of local optical modes, the stationary optical amplitudes are $\bar{c}_j = \sqrt{I_{\rm sat}\left(\frac{g}{\gamma}-1\right)}\,e^{-i\left(jl+\varphi\right)}.$

We linearize the full Langevin equations (\ref{full eq}) around the stationary mean-field solution $c_j = \left(\bar{c}_j +\delta c_j\right)e^{-i\Omega t}$ obtaining the linearized Langevin equations (\ref{bog eq}), where $\Omega = \omega_l$. The dynamical matrix $\mathcal{D}$ is given by Eq.~(\ref{dyn m}), where $\Gamma_{jj} =  -\gamma + \frac{\gamma^2}{g}= - \bar{\gamma}$, $\Delta_{jj} =  -  \bar{\gamma} e^{-2i\left(jl+\varphi\right)}$, and $\Gamma_{jk}=\Delta_{jk}=0$ for $j\neq k$.
Switching to the basis of passive-system normal modes, we diagonalize the Hermitian part $\mathcal{H}$ of the dynamical matrix. Since the passive-system normal modes are plain waves, the matrix elements of the anti-Hermitian part $\tilde{\mathcal{A}}$ can be explicitly evaluated, $\tilde{\Gamma}_{mm} = - \bar{\gamma}$, $\tilde{\Delta}_{m(2l-m)} = - \bar{\gamma} e^{-2i\varphi}$, and $\tilde{\Gamma}_{mn}=\tilde{\Delta}_{m(2l-n)}=0$ for $m\neq n$. The diagonal block $\tilde{\mathbf{\Gamma}}$ describes decay of the normal modes at rate $\bar{\gamma}$. The off-diagonal block $\tilde{\mathbf{\Delta}}$ describes the coupling of passive-system normal modes whose quasi-momenta $m$ and $n$ satisfy the condition $m+n = 2l$. Since this condition is satisfied only for mode pairs $(m,2l-m)$, the coupling of passive-system normal modes is exactly described by the $2\times2$ dynamical matrix $\tilde{\mathcal{D}}^{(m,2l-m)}$ given by Eq.~(\ref{dyn 2}). The complex spectrum of elementary excitations is exactly determined by the eigenvalues of the $2\times2$ dynamical matrix
\begin{equation}\label{hyb eps}
\epsilon_{\pm}^{(m,2l-m)} = \bar{\omega}_{m(2l-m)} - i \bar{\gamma} \pm \frac{1}{2}\sqrt{\delta\omega_{m(2l-m)}^2 - 4\bar{\gamma}^2},
\end{equation}
$\bar{\omega}_{m(2l-m)} = \tfrac{1}{2}(\omega_m - \omega_{(2l-m)})$,
$\delta\omega_{m(2l-m)} =  \omega_m + \omega_{2l-m} -2\Omega$.

The spectrum of elementary excitations in the one-dimensional laser array depends only on the dispersion of the passive system frequencies  $\omega_m$. If $|\delta\omega_{m(2l-m)}|>0$ for all $m\neq l$, lasing of a single mode $l$ is a stable steady state as all ${\rm Im}\,\epsilon_{\pm}^{(m,2l-m)} <0$ except from the non-decaying mode $\epsilon_+^{(l,l)}=0$. We can expand the dispersion of passive-system frequencies $\omega_m = \Omega + v_1 \left( m - l \right) + v_2 \left(m-l\right)^2 + \mathcal{O}\left(\left(m-l\right)^3\right)$ around the lasing frequency $\Omega$. If higher order terms in the expansion are negligible, the detuning between modes $m$ and $2l-m$ is $\delta \omega_{m(2l-m)} = 2 v_2\left(m-l\right)^2 + \mathcal{O}\left(\left(m-l\right)^4\right)$. If the nonlinear coefficient $v_2$ is sufficiently small such that $|v_2|(m-l)^2\ll\bar{\gamma}$ for $m$ close to $l$, the complex frequencies are $\epsilon_+^{(m,2l-m)}\approx \bar{\omega}_m - i\frac{v_2^2}{2\bar{\gamma}}\left(m-l\right)^4$ and $\epsilon_-^{(m,2l-m)}\approx \bar{\omega}_m - i\left[2\bar{\gamma} - \frac{v_2^2}{2\bar{\gamma}}\left(m-l\right)^4\right]$ corresponding to slowly-decaying modes and fast-decaying modes, respectively. For a small nonlinear coefficient $v_2$, the decay rate of the slowly-decaying modes can be orders of magnitude smaller than any other energy scale in the system $g,\gamma,\omega_m$, leading to long-lived elementary excitations.

Long-lived elementary excitations generically appear in one-dimensional laser arrays if the dispersion of the passive-system frequencies $\omega_m$ is linear around the lasing frequency.

\section*{Appendix E: Correlations of amplitude and phase fluctuations }
\label{app phi0}
\setcounter{equation}{0}
\renewcommand{\theequation}{E{\arabic{equation}}}
Here, we provide details about how the optical spectrum and the second-order autocorrelation function are derived and how they are related to the normal modes of elementary excitations.

It is convenient to study noise in terms of amplitude and phase fluctuations, due to the $U(1)$ symmetry of the mean-field dynamical equations, $c_j\rightarrow c_j \,e^{i\varphi}$, where $\varphi$ is an arbitrary overall phase. The coherence properties of a laser driven well above threshold are directly determined by the correlations in amplitude fluctuations and phase fluctuations \cite{gardiner2004}. Amplitude fluctuations $\delta C_j$ and phase fluctuations $\delta \theta_j$ are linearly related to the fluctuations of complex amplitudes $ \delta c_j $ and $ \delta c_j^* $
\begin{gather}\label{bog to amp}
\delta C_j=\frac{e^{-i\bar{\theta}_j} \delta c_j + e^{i\bar{\theta}_j} \delta c_j^* }{2},\,\,\,\,
\,\delta\theta_j=\frac{e^{-i\bar{\theta}_j} \delta c_j - e^{i\bar{\theta}_j} \delta c_j^* }{2i\bar{C}_j},
\end{gather}
where $\bar{c}_j = \bar{C}_j e^{i\bar{\theta}_j}$. This relation can be described by the linear transformation
\begin{equation}
\begin{pmatrix}
\delta\mathbf{ C} \\
\delta \mathbf{\Theta}
\end{pmatrix}
=
\mathcal{T}\begin{pmatrix}
\delta\mathbf{ c} \\
\delta\mathbf{ c}^*
\end{pmatrix},
\end{equation}
where $\delta\Theta_j = \bar{C}_j\,\delta\theta_j $. The linearized Langevin equations around the mean-field steady-state for the amplitude and phase fluctuations are
\begin{equation}
i\frac{\rm d}{{\rm d} t}
\begin{pmatrix}
\delta\mathbf{ C} \\
\delta\mathbf{ \Theta}
\end{pmatrix}
=
\mathcal{T}\mathbf{\mathcal{D}}\mathcal{T}^{-1}
\begin{pmatrix}
\delta\mathbf{ C} \\
\delta\mathbf{ \Theta}
\end{pmatrix}
+
\frac{i}{\sqrt{2}}\mathcal{Q}
\begin{pmatrix}
\mathbf{C}_{\rm in} \\
\mathbf{\Theta}_{\rm in}
\end{pmatrix},
\end{equation}
where $\mathbf{C}_{\rm in}$ and $\mathbf{\Theta}_{\rm in}$ describe real-valued Gaussian white noise, with following correlations $\langle C_{j,{\rm in}}(t)C_{k,{\rm in}}(t')\rangle = \delta_{jk}\delta(t-t')$, $\langle \Theta_{j,{\rm in}}(t)\Theta_{k,{\rm in}}(t')\rangle = \delta_{jk}\delta(t-t')$, and $\langle \bar{C}_{j,{\rm in}}(t)\Theta_{k,{\rm in}}(t')\rangle =0$. Amplitude and phase fluctuations are linearly related to the normal modes of elementary excitations
\begin{equation}
\begin{pmatrix}
\delta\mathbf{ C} \\
\delta \mathbf{\Theta}
\end{pmatrix}
=
\mathcal{W}\mathcal{N},
\end{equation}
where $\mathcal{W} = \mathcal{V}\mathcal{T}$,  columns of the matrix $\mathcal{V}$ are the normal modes of elementary excitations $\mathcal{E}^{(\alpha)}$ described in Sec.~\ref{sec elm} and the vector $\mathcal{N}$ contains the complex amplitudes of these normal modes. Non-equal-time phase and amplitude autocorrelations can be expressed in terms of normal modes' correlations
\begin{gather}
\langle \delta C_j (t) \delta C_j(t +\Delta t) \rangle =\sum_{\alpha,\beta= 1}^{2N} \mathcal{W}_{j\alpha} \langle \mathcal{N}_{\alpha}(t)\mathcal{N}_{\beta}^{*}(t + \Delta t)\rangle \mathcal{W}^{\dagger}_{\beta j},\\
\langle \left[\delta \theta_j (t) - \delta \theta_j(t+\Delta t) \right]^2\rangle \nonumber\\
=\frac{1}{\bar{C}_j^2}\sum_{\alpha,\beta = 1}^{2N} \mathcal{W}_{(j+N)\alpha} \langle | \mathcal{N}_{\alpha}(t) - \mathcal{N}_{\beta}(t + \Delta t)|^2\rangle \mathcal{W}^{\dagger}_{\beta(j+N)}.\label{corr th}
\end{gather}
The correlations of normal modes are
\begin{gather}
\langle \mathcal{N}_{\alpha}(t)\mathcal{N}_{\beta}^{*}(t+ \Delta t)\rangle = \frac{i\left(\mathcal{R}\mathcal{R}^{\dagger}\right)_{\alpha\beta}}{ \left(\epsilon^{(\beta)}\right)^* - \epsilon^{(\alpha)} }\,e^{i\epsilon^{(\alpha)}\Delta t},\,\,\Delta t<0,\label{corr 1}\\
\langle \mathcal{N}_{\alpha}(t)\mathcal{N}_{\beta}^{*}(t + \Delta t)\rangle = \frac{i\left(\mathcal{R}\mathcal{R}^{\dagger}\right)_{\alpha\beta}}{\left(\epsilon^{(\beta)}\right)^*-\epsilon^{(\alpha)} }\,e^{i\left(\epsilon^{(\beta)}\right)^*\Delta t},\,\,\Delta t > 0,\label{corr 2}
\end{gather}
for all normal modes $\alpha$ and $\beta$ except from the autocorrelation of the non-decaying mode, i.e.~for $\alpha =\beta=\eta$ and $\epsilon^{(\eta)}=0$. Note that the non-decaying mode is related only to phase fluctuations. As a result, the relevant autocorrelation of the non-decaying mode is
\begin{gather}\label{corr g}
\langle | \mathcal{N}_{\eta}(t) - \mathcal{N}_{\eta}(t+\Delta t)|^2\rangle =\left(\mathcal{R}\mathcal{R}^{\dagger}\right)_{\eta\eta}\,|\Delta t|.
\end{gather}

The dominant contribution in the autocorrelation of complex optical amplitudes reads
\begin{equation}
 \langle c_j(t) c_j^*(t+\Delta t)\rangle \approx \bar{C}_j^{2}\,e^{i\Omega\Delta t} \,e^{-\langle \left[\delta \theta_j (t) - \delta \theta_j(t + \Delta t) \right]^2\rangle/2},
\end{equation}
where amplitude fluctuations are neglected, since they are small in comparison to the large mean-field occupation $\bar{C}_j^{2}$  \cite{gardiner2004}. Using Eq.~(\ref{corr th}), we express $\langle \left[\delta \theta_j (t) - \delta \theta_j(t + \Delta t) \right]^2\rangle$ in terms of correlations in the normal modes of elementary excitations (\ref{corr 1}), (\ref{corr 2}) and (\ref{corr g}) to derive the optical spectrum Eq.~(\ref{spec ee}) in the main text, where we neglect correlations between different normal modes $\langle | \mathcal{N}_{\alpha}(t) - \mathcal{N}_{\beta}(t + \Delta t)|^2\rangle$ for $\alpha\neq\beta$. Only long-lived elementary excitations have significant contribution to the optical spectrum due to their large occupation $n_{\alpha}$. Since the corresponding normal modes are formed from edge modes, they have a large detuning in the real parts of complex frequencies, which suppresses the correlations between different normal modes $\langle | \mathcal{N}_{\alpha}(t) - \mathcal{N}_{\beta}(t + \Delta t)|^2\rangle$ for $\alpha\neq\beta$.

Similarly, we can express the second-order autocorrelation function (\ref{g2 eq}) in terms of correlations in the normal modes of elementary excitations. 

\section*{Appendix F: Disorder}
\label{disorder}
\setcounter{equation}{0}
\setcounter{figure}{0}
\renewcommand{\theequation}{F{\arabic{equation}}}
\renewcommand{\thefigure}{F{\arabic{figure}}}
\begin{figure}[h]
\centering
\includegraphics[width=\linewidth]{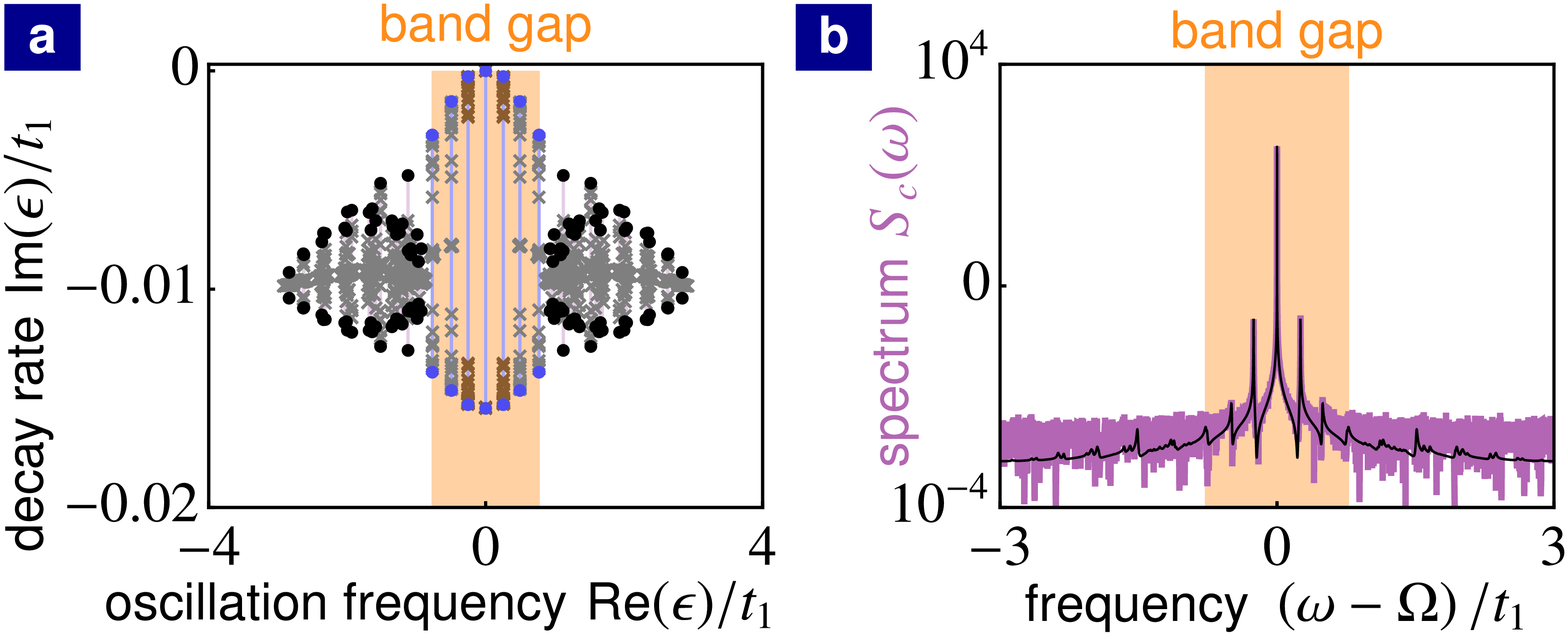}
\caption{Effects of moderate disorder. (a) Complex spectrum of elementary excitations for 30 disorder realizations (brown and gray crosses) for $\Omega/t_1\approx0$ compared to complex spectrum of elementary excitations without disorder with band gap (orange region), bulk modes (black points) as well as edge modes (blue points) for the lasing frequency $\Omega/t_1 = 0$. Purple and blue lines show the splitting in imaginary parts of complex frequencies due to the hybridization of bulk modes and edge modes, respectively. (b) Optical spectrum of a pumped optical site lying at the edge of the topological array for a single disorder realization and for lasing at the frequency $\Omega/t_1 =0.01$. Linearization of Langevin equations around the mean-field steady state (black line) and numerical simulations of non-linear Langevin equations (purple line). The orange region shows the band gap in the spectrum of elementary excitations. (Parameters: (a) and (b) $t_2/t_1=0.15$, $\phi=\pi/2$, $\gamma/t_1=0.01$, $g/t_1=0.05$, $\sigma/t_1=0.1$; (b) $I_{\rm sat}\gamma/q=25$, $\Omega = 0.01$)
}
\label{fig dis}
\end{figure}

Here we study effects of a moderate on-site disorder on the spectrum of elementary excitations and long-lived elementary excitations. We consider lasing at a frequency, which lies in the middle of the passive-system band gap. The on-site disorder is represented by a Gaussian distribution of on-site frequencies $\nu_j$ with a zero mean value and a standard deviation $\sigma$. We consider a moderate disorder with the standard deviation $\sigma$ smaller than the size of the passive-system band gap $6\sqrt{3}t_2\sin\phi$ but larger than the decay rate $\gamma$ and gain $g$.

We compare the spectrum of elementary excitations for 30 disorder realizations and the spectrum of elementary excitations without disorder in Fig.~\ref{fig dis}a. The large hybridization of edge modes with frequencies close to the lasing frequency leads to long-lived elementary excitations with a very small decay rate for all disorder realizations (brown crosses). This shows the robustness of long-lived elementary excitations against disorder. The incoherent occupation $n_{\alpha}$ of the corresponding slowly-decaying normal modes is large even in the presence of moderate on-site disorder and it leads to satellite peaks in the optical spectrum, see Fig.~\ref{fig dis}b.

\section*{Appendix G: Long-lived elementary excitations in a trivial laser}
\label{trivial}
\setcounter{equation}{0}
\setcounter{figure}{0}
\renewcommand{\theequation}{G{\arabic{equation}}}
\renewcommand{\thefigure}{G{\arabic{figure}}}
\begin{figure}[h]
\centering
\includegraphics[width=\linewidth]{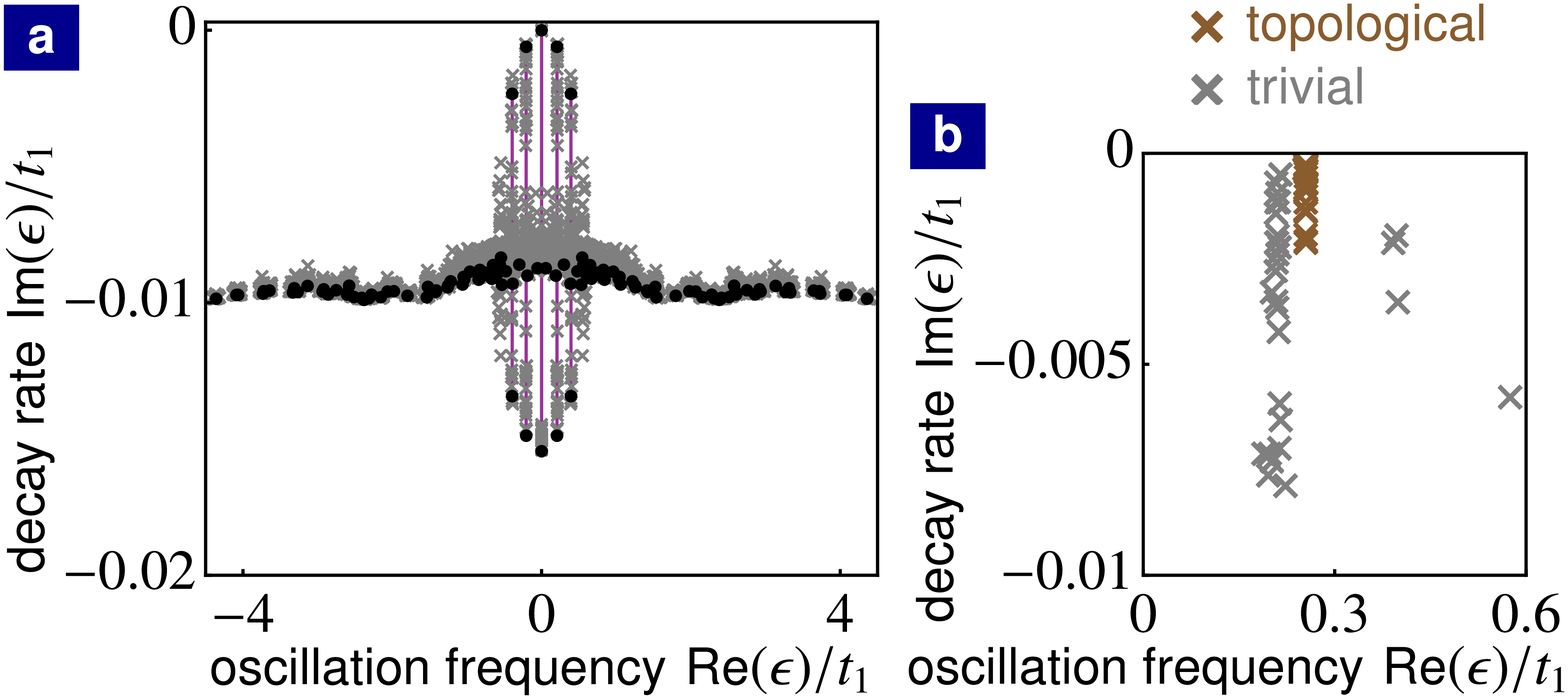}
\caption{Effects of moderate disorder in a trivial laser based on the Haldane model compared to a topological laser. (a) Complex spectrum of elementary excitations in a trivial laser for 30 disorder realizations (gray crosses) and for the lasing frequency $\Omega/t_1\approx1$ compared to complex spectrum of elementary excitations without disorder (black points) for $\Omega/t_1 = 1.03$. Purple lines show the splitting in the imaginary parts of complex frequencies due to the hybridization of trivial edge modes. (b) The complex frequency with the smallest imaginary part $\textrm{min}_{\alpha\neq\eta}|\textrm{Im}\epsilon^{(\alpha)}|$ for each disorder realization (30 in total), for a trivial laser with $\Omega/t_1 \approx 1$ (gray crosses) as well as for the topological laser with $\Omega\approx0$ (brown crosses). (Parameters: (a) and (b) $t_2/t_1=0.15$, $\gamma/t_1=0.01$, $g/t_1=0.05$, $\sigma/t_1=0.1$; (a) and (b) gray crosses $\phi=2\pi/3$, $M/t_1=0.8$; (b) brown crosses $\phi=\pi/2$, $M/t_1=0$ )
}
\label{trivial}
\end{figure}

We now study long-lived elementary excitations in a trivial laser based on the Haldane model. We focus on the effects of disorder on the long-lived elementary excitations and we compare them to the effects of disorder on long-lived elementary excitations in a topological laser studied in Appendix~F.

We introduce the Haldane mass term $M$ in the Hamiltonian
\begin{equation}
\hat{H} = M \sum_{j} \mu_j \hat{c}^{\dagger}_j \hat{c}_j +   t_1\sum_{\rm n.n.} \hat{c}^{\dagger}_j \hat{c}_k + t_2\sum_{\rm n.n.n.} e^{i\phi_{jk}} \hat{c}^{\dagger}_j \hat{c}_k,
\end{equation}
which is a staggered on-site potential with $\mu_j = 1$ for sites in sublattice A and $\mu_j = -1$ for sites in sublattice B \cite{haldane1988}. For $|M| > 3\sqrt{3}t_2 |\sin\phi |$, a trivial band gap opens. For this topologically-trivial phase, the finite-size array depicted in Fig.~\ref{steady}a has trivial edge modes, whose energies lie within the bulk bands. On the mean-field level, we observe single-mode lasing of a trivial edge mode. We plot the spectrum of elementary excitations (black points) in Fig.~\ref{trivial}a for the lasing frequency $\Omega/t_1 = 1.03$. For this lasing frequency, long-lived elementary excitations whose decay rates are one order of magnitude smaller than any other energy scale in the system ($\gamma$, $g$, $t_1$ and $t_2$) appear due to a large hybridization of trivial edge modes. This is analogous to the long-lived elementary excitations in a topological laser due to the hybridization of topological edge modes discussed in Sec.~\ref{llee}.

For moderate on-site disorder, the single-mode lasing is still stable. We compare the spectrum of elementary excitations (gray crosses) for 30 disorder realizations and the spectrum of elementary excitations without disorder (black points) in Fig.~\ref{trivial}a. The decay rate of elementary excitations with small oscillation frequencies varies strongly in each disorder realization. For some disorder realizations, the hybridization of trivial edge modes is obstructed and the decay rate of the corresponding elementary excitations increases by one order of magnitude.
This can be seen in Fig.~\ref{trivial}b where we plot the complex frequency (gray crosses) with the smallest imaginary part $\textrm{min}_{\alpha\neq\eta}|\textrm{Im}\epsilon^{(\alpha)}|$ for each disorder realization.
In contrast to the long-lived elementary excitations in a trivial laser, the long-lived elementary excitations in a topological laser are robust against disorder, see Appendix~F. In the topological laser, the smallest decay rate of elementary excitations only moderately changes depending on each disorder realization and the corresponding oscillation frequency is unchanged, see brown crosses in Fig.~\ref{trivial}b.

\section*{Acknowledgment}

We thank D. Malz for insightful discussions. This work was supported by the European Union’s Horizon 2020 research and innovation programme under grant agreement No 732894 (FET Proactive HOT). A.N. holds a University Research Fellowship from the Royal Society and acknowledges support from the Winton Programme for the Physics of Sustainability.

\section*{Note added}

During the final stage of this project, a preprint \cite{amelio2019} appeared, investigating coherence properties of topological lasers. Their numerical analysis of topological lasing in arrays of large sizes complements our study of topological lasing reported here.

\bibliographystyle{apsrev4-1}
\bibliography{TL_paper}

\begin{thebibliography}{32}%
\makeatletter
\providecommand \@ifxundefined [1]{%
 \@ifx{#1\undefined}
}%
\providecommand \@ifnum [1]{%
 \ifnum #1\expandafter \@firstoftwo
 \else \expandafter \@secondoftwo
 \fi
}%
\providecommand \@ifx [1]{%
 \ifx #1\expandafter \@firstoftwo
 \else \expandafter \@secondoftwo
 \fi
}%
\providecommand \natexlab [1]{#1}%
\providecommand \enquote  [1]{``#1''}%
\providecommand \bibnamefont  [1]{#1}%
\providecommand \bibfnamefont [1]{#1}%
\providecommand \citenamefont [1]{#1}%
\providecommand \href@noop [0]{\@secondoftwo}%
\providecommand \href [0]{\begingroup \@sanitize@url \@href}%
\providecommand \@href[1]{\@@startlink{#1}\@@href}%
\providecommand \@@href[1]{\endgroup#1\@@endlink}%
\providecommand \@sanitize@url [0]{\catcode `\\12\catcode `\$12\catcode
  `\&12\catcode `\#12\catcode `\^12\catcode `\_12\catcode `\%12\relax}%
\providecommand \@@startlink[1]{}%
\providecommand \@@endlink[0]{}%
\providecommand \url  [0]{\begingroup\@sanitize@url \@url }%
\providecommand \@url [1]{\endgroup\@href {#1}{\urlprefix }}%
\providecommand \urlprefix  [0]{URL }%
\providecommand \Eprint [0]{\href }%
\providecommand \doibase [0]{http://dx.doi.org/}%
\providecommand \selectlanguage [0]{\@gobble}%
\providecommand \bibinfo  [0]{\@secondoftwo}%
\providecommand \bibfield  [0]{\@secondoftwo}%
\providecommand \translation [1]{[#1]}%
\providecommand \BibitemOpen [0]{}%
\providecommand \bibitemStop [0]{}%
\providecommand \bibitemNoStop [0]{.\EOS\space}%
\providecommand \EOS [0]{\spacefactor3000\relax}%
\providecommand \BibitemShut  [1]{\csname bibitem#1\endcsname}%
\let\auto@bib@innerbib\@empty
\bibitem [{\citenamefont {Ozawa}\ \emph {et~al.}(2019)\citenamefont {Ozawa},
  \citenamefont {Price}, \citenamefont {Amo}, \citenamefont {Goldman},
  \citenamefont {Hafezi}, \citenamefont {Lu}, \citenamefont {Rechtsman},
  \citenamefont {Schuster}, \citenamefont {Simon}, \citenamefont {Zilberberg},\
  and\ \citenamefont {Carusotto}}]{ozawa2019}%
  \BibitemOpen
  \bibfield  {author} {\bibinfo {author} {\bibfnamefont {T.}~\bibnamefont
  {Ozawa}}, \bibinfo {author} {\bibfnamefont {H.~M.}\ \bibnamefont {Price}},
  \bibinfo {author} {\bibfnamefont {A.}~\bibnamefont {Amo}}, \bibinfo {author}
  {\bibfnamefont {N.}~\bibnamefont {Goldman}}, \bibinfo {author} {\bibfnamefont
  {M.}~\bibnamefont {Hafezi}}, \bibinfo {author} {\bibfnamefont
  {L.}~\bibnamefont {Lu}}, \bibinfo {author} {\bibfnamefont {M.~C.}\
  \bibnamefont {Rechtsman}}, \bibinfo {author} {\bibfnamefont {D.}~\bibnamefont
  {Schuster}}, \bibinfo {author} {\bibfnamefont {J.}~\bibnamefont {Simon}},
  \bibinfo {author} {\bibfnamefont {O.}~\bibnamefont {Zilberberg}}, \ and\
  \bibinfo {author} {\bibfnamefont {I.}~\bibnamefont {Carusotto}},\ }\href
  {\doibase 10.1103/RevModPhys.91.015006} {\bibfield  {journal} {\bibinfo
  {journal} {Rev. Mod. Phys.}\ }\textbf {\bibinfo {volume} {91}},\ \bibinfo
  {pages} {015006} (\bibinfo {year} {2019})}\BibitemShut {NoStop}%
\bibitem [{\citenamefont {Martinez~Alvarez}\ \emph {et~al.}(2018)\citenamefont
  {Martinez~Alvarez}, \citenamefont {Barrios~Vargas}, \citenamefont
  {Berdakin},\ and\ \citenamefont {Foa~Torres}}]{martinezalvarez2018}%
  \BibitemOpen
  \bibfield  {author} {\bibinfo {author} {\bibfnamefont {V.~M.}\ \bibnamefont
  {Martinez~Alvarez}}, \bibinfo {author} {\bibfnamefont {J.~E.}\ \bibnamefont
  {Barrios~Vargas}}, \bibinfo {author} {\bibfnamefont {M.}~\bibnamefont
  {Berdakin}}, \ and\ \bibinfo {author} {\bibfnamefont {L.~E.~F.}\ \bibnamefont
  {Foa~Torres}},\ }\href {\doibase 10.1140/epjst/e2018-800091-5} {\bibfield
  {journal} {\bibinfo  {journal} {Eur. Phys. J. Spec. Top.}\ }\textbf {\bibinfo
  {volume} {227}},\ \bibinfo {pages} {1295} (\bibinfo {year}
  {2018})}\BibitemShut {NoStop}%
\bibitem [{\citenamefont {Zhao}\ \emph {et~al.}(2019)\citenamefont {Zhao},
  \citenamefont {Qiao}, \citenamefont {Wu}, \citenamefont {Midya},
  \citenamefont {Longhi},\ and\ \citenamefont {Feng}}]{zhao2019}%
  \BibitemOpen
  \bibfield  {author} {\bibinfo {author} {\bibfnamefont {H.}~\bibnamefont
  {Zhao}}, \bibinfo {author} {\bibfnamefont {X.}~\bibnamefont {Qiao}}, \bibinfo
  {author} {\bibfnamefont {T.}~\bibnamefont {Wu}}, \bibinfo {author}
  {\bibfnamefont {B.}~\bibnamefont {Midya}}, \bibinfo {author} {\bibfnamefont
  {S.}~\bibnamefont {Longhi}}, \ and\ \bibinfo {author} {\bibfnamefont
  {L.}~\bibnamefont {Feng}},\ }\href {\doibase 10.1126/science.aay1064}
  {\bibfield  {journal} {\bibinfo  {journal} {Science}\ }\textbf {\bibinfo
  {volume} {365}},\ \bibinfo {pages} {1163} (\bibinfo {year}
  {2019})}\BibitemShut {NoStop}%
\bibitem [{\citenamefont {Cerjan}\ \emph {et~al.}(2019)\citenamefont {Cerjan},
  \citenamefont {Huang}, \citenamefont {Wang}, \citenamefont {Chen},
  \citenamefont {Chong},\ and\ \citenamefont {Rechtsman}}]{cerjan2019}%
  \BibitemOpen
  \bibfield  {author} {\bibinfo {author} {\bibfnamefont {A.}~\bibnamefont
  {Cerjan}}, \bibinfo {author} {\bibfnamefont {S.}~\bibnamefont {Huang}},
  \bibinfo {author} {\bibfnamefont {M.}~\bibnamefont {Wang}}, \bibinfo {author}
  {\bibfnamefont {K.~P.}\ \bibnamefont {Chen}}, \bibinfo {author}
  {\bibfnamefont {Y.}~\bibnamefont {Chong}}, \ and\ \bibinfo {author}
  {\bibfnamefont {M.~C.}\ \bibnamefont {Rechtsman}},\ }\href {\doibase
  10.1038/s41566-019-0453-z} {\bibfield  {journal} {\bibinfo  {journal} {Nat.
  Photonics}\ }\textbf {\bibinfo {volume} {13}},\ \bibinfo {pages} {623}
  (\bibinfo {year} {2019})}\BibitemShut {NoStop}%
\bibitem [{\citenamefont {Yuan}\ \emph {et~al.}(2016)\citenamefont {Yuan},
  \citenamefont {Shi},\ and\ \citenamefont {Fan}}]{yuan2016}%
  \BibitemOpen
  \bibfield  {author} {\bibinfo {author} {\bibfnamefont {L.}~\bibnamefont
  {Yuan}}, \bibinfo {author} {\bibfnamefont {Y.}~\bibnamefont {Shi}}, \ and\
  \bibinfo {author} {\bibfnamefont {S.}~\bibnamefont {Fan}},\ }\href {\doibase
  10.1364/OL.41.000741} {\bibfield  {journal} {\bibinfo  {journal} {Opt.
  Lett.}\ }\textbf {\bibinfo {volume} {41}},\ \bibinfo {pages} {741} (\bibinfo
  {year} {2016})}\BibitemShut {NoStop}%
\bibitem [{\citenamefont {Ozawa}\ \emph {et~al.}(2016)\citenamefont {Ozawa},
  \citenamefont {Price}, \citenamefont {Goldman}, \citenamefont {Zilberberg},\
  and\ \citenamefont {Carusotto}}]{ozawa2016}%
  \BibitemOpen
  \bibfield  {author} {\bibinfo {author} {\bibfnamefont {T.}~\bibnamefont
  {Ozawa}}, \bibinfo {author} {\bibfnamefont {H.~M.}\ \bibnamefont {Price}},
  \bibinfo {author} {\bibfnamefont {N.}~\bibnamefont {Goldman}}, \bibinfo
  {author} {\bibfnamefont {O.}~\bibnamefont {Zilberberg}}, \ and\ \bibinfo
  {author} {\bibfnamefont {I.}~\bibnamefont {Carusotto}},\ }\href {\doibase
  10.1103/PhysRevA.93.043827} {\bibfield  {journal} {\bibinfo  {journal} {Phys.
  Rev. A}\ }\textbf {\bibinfo {volume} {93}},\ \bibinfo {pages} {043827}
  (\bibinfo {year} {2016})}\BibitemShut {NoStop}%
\bibitem [{\citenamefont {Lustig}\ \emph {et~al.}(2019)\citenamefont {Lustig},
  \citenamefont {Weimann}, \citenamefont {Plotnik}, \citenamefont {Lumer},
  \citenamefont {Bandres}, \citenamefont {Szameit},\ and\ \citenamefont
  {Segev}}]{lustig2019}%
  \BibitemOpen
  \bibfield  {author} {\bibinfo {author} {\bibfnamefont {E.}~\bibnamefont
  {Lustig}}, \bibinfo {author} {\bibfnamefont {S.}~\bibnamefont {Weimann}},
  \bibinfo {author} {\bibfnamefont {Y.}~\bibnamefont {Plotnik}}, \bibinfo
  {author} {\bibfnamefont {Y.}~\bibnamefont {Lumer}}, \bibinfo {author}
  {\bibfnamefont {M.~A.}\ \bibnamefont {Bandres}}, \bibinfo {author}
  {\bibfnamefont {A.}~\bibnamefont {Szameit}}, \ and\ \bibinfo {author}
  {\bibfnamefont {M.}~\bibnamefont {Segev}},\ }\href {\doibase
  10.1038/s41586-019-0943-7} {\bibfield  {journal} {\bibinfo  {journal}
  {Nature}\ }\textbf {\bibinfo {volume} {567}},\ \bibinfo {pages} {356}
  (\bibinfo {year} {2019})}\BibitemShut {NoStop}%
\bibitem [{\citenamefont {Dutt}\ \emph {et~al.}(2020)\citenamefont {Dutt},
  \citenamefont {Lin}, \citenamefont {Yuan}, \citenamefont {Minkov},
  \citenamefont {Xiao},\ and\ \citenamefont {Fan}}]{dutt2020}%
  \BibitemOpen
  \bibfield  {author} {\bibinfo {author} {\bibfnamefont {A.}~\bibnamefont
  {Dutt}}, \bibinfo {author} {\bibfnamefont {Q.}~\bibnamefont {Lin}}, \bibinfo
  {author} {\bibfnamefont {L.}~\bibnamefont {Yuan}}, \bibinfo {author}
  {\bibfnamefont {M.}~\bibnamefont {Minkov}}, \bibinfo {author} {\bibfnamefont
  {M.}~\bibnamefont {Xiao}}, \ and\ \bibinfo {author} {\bibfnamefont
  {S.}~\bibnamefont {Fan}},\ }\href {\doibase 10.1126/science.aaz3071}
  {\bibfield  {journal} {\bibinfo  {journal} {Science}\ }\textbf {\bibinfo
  {volume} {367}},\ \bibinfo {pages} {59} (\bibinfo {year} {2020})}\BibitemShut
  {NoStop}%
\bibitem [{\citenamefont {Smirnova}\ \emph {et~al.}(2020)\citenamefont
  {Smirnova}, \citenamefont {Leykam}, \citenamefont {Chong},\ and\
  \citenamefont {Kivshar}}]{smirnova2020}%
  \BibitemOpen
  \bibfield  {author} {\bibinfo {author} {\bibfnamefont {D.}~\bibnamefont
  {Smirnova}}, \bibinfo {author} {\bibfnamefont {D.}~\bibnamefont {Leykam}},
  \bibinfo {author} {\bibfnamefont {Y.}~\bibnamefont {Chong}}, \ and\ \bibinfo
  {author} {\bibfnamefont {Y.}~\bibnamefont {Kivshar}},\ }\href {\doibase
  10.1063/1.5142397} {\bibfield  {journal} {\bibinfo  {journal} {Appl. Phys.
  Rev.}\ }\textbf {\bibinfo {volume} {7}},\ \bibinfo {pages} {021306} (\bibinfo
  {year} {2020})}\BibitemShut {NoStop}%
\bibitem [{\citenamefont {{St-Jean}}\ \emph {et~al.}(2017)\citenamefont
  {{St-Jean}}, \citenamefont {Goblot}, \citenamefont {Galopin}, \citenamefont
  {Lema{\^i}tre}, \citenamefont {Ozawa}, \citenamefont {Gratiet}, \citenamefont
  {Sagnes}, \citenamefont {Bloch},\ and\ \citenamefont {Amo}}]{st-jean2017}%
  \BibitemOpen
  \bibfield  {author} {\bibinfo {author} {\bibfnamefont {P.}~\bibnamefont
  {{St-Jean}}}, \bibinfo {author} {\bibfnamefont {V.}~\bibnamefont {Goblot}},
  \bibinfo {author} {\bibfnamefont {E.}~\bibnamefont {Galopin}}, \bibinfo
  {author} {\bibfnamefont {A.}~\bibnamefont {Lema{\^i}tre}}, \bibinfo {author}
  {\bibfnamefont {T.}~\bibnamefont {Ozawa}}, \bibinfo {author} {\bibfnamefont
  {L.~L.}\ \bibnamefont {Gratiet}}, \bibinfo {author} {\bibfnamefont
  {I.}~\bibnamefont {Sagnes}}, \bibinfo {author} {\bibfnamefont
  {J.}~\bibnamefont {Bloch}}, \ and\ \bibinfo {author} {\bibfnamefont
  {A.}~\bibnamefont {Amo}},\ }\href {\doibase 10.1038/s41566-017-0006-2}
  {\bibfield  {journal} {\bibinfo  {journal} {Nat. Photonics}\ }\textbf
  {\bibinfo {volume} {11}},\ \bibinfo {pages} {651} (\bibinfo {year}
  {2017})}\BibitemShut {NoStop}%
\bibitem [{\citenamefont {Parto}\ \emph {et~al.}(2018)\citenamefont {Parto},
  \citenamefont {Wittek}, \citenamefont {Hodaei}, \citenamefont {Harari},
  \citenamefont {Bandres}, \citenamefont {Ren}, \citenamefont {Rechtsman},
  \citenamefont {Segev}, \citenamefont {Christodoulides},\ and\ \citenamefont
  {Khajavikhan}}]{parto2018}%
  \BibitemOpen
  \bibfield  {author} {\bibinfo {author} {\bibfnamefont {M.}~\bibnamefont
  {Parto}}, \bibinfo {author} {\bibfnamefont {S.}~\bibnamefont {Wittek}},
  \bibinfo {author} {\bibfnamefont {H.}~\bibnamefont {Hodaei}}, \bibinfo
  {author} {\bibfnamefont {G.}~\bibnamefont {Harari}}, \bibinfo {author}
  {\bibfnamefont {M.~A.}\ \bibnamefont {Bandres}}, \bibinfo {author}
  {\bibfnamefont {J.}~\bibnamefont {Ren}}, \bibinfo {author} {\bibfnamefont
  {M.~C.}\ \bibnamefont {Rechtsman}}, \bibinfo {author} {\bibfnamefont
  {M.}~\bibnamefont {Segev}}, \bibinfo {author} {\bibfnamefont {D.~N.}\
  \bibnamefont {Christodoulides}}, \ and\ \bibinfo {author} {\bibfnamefont
  {M.}~\bibnamefont {Khajavikhan}},\ }\href {\doibase
  10.1103/PhysRevLett.120.113901} {\bibfield  {journal} {\bibinfo  {journal}
  {Phys. Rev. Lett.}\ }\textbf {\bibinfo {volume} {120}},\ \bibinfo {pages}
  {113901} (\bibinfo {year} {2018})}\BibitemShut {NoStop}%
\bibitem [{\citenamefont {Zhao}\ \emph {et~al.}(2018)\citenamefont {Zhao},
  \citenamefont {Miao}, \citenamefont {Teimourpour}, \citenamefont {Malzard},
  \citenamefont {{El-Ganainy}}, \citenamefont {Schomerus},\ and\ \citenamefont
  {Feng}}]{zhao2018}%
  \BibitemOpen
  \bibfield  {author} {\bibinfo {author} {\bibfnamefont {H.}~\bibnamefont
  {Zhao}}, \bibinfo {author} {\bibfnamefont {P.}~\bibnamefont {Miao}}, \bibinfo
  {author} {\bibfnamefont {M.~H.}\ \bibnamefont {Teimourpour}}, \bibinfo
  {author} {\bibfnamefont {S.}~\bibnamefont {Malzard}}, \bibinfo {author}
  {\bibfnamefont {R.}~\bibnamefont {{El-Ganainy}}}, \bibinfo {author}
  {\bibfnamefont {H.}~\bibnamefont {Schomerus}}, \ and\ \bibinfo {author}
  {\bibfnamefont {L.}~\bibnamefont {Feng}},\ }\href {\doibase
  10.1038/s41467-018-03434-2} {\bibfield  {journal} {\bibinfo  {journal} {Nat.
  Commun.}\ }\textbf {\bibinfo {volume} {9}},\ \bibinfo {pages} {981} (\bibinfo
  {year} {2018})}\BibitemShut {NoStop}%
\bibitem [{\citenamefont {Bahari}\ \emph {et~al.}(2017)\citenamefont {Bahari},
  \citenamefont {Ndao}, \citenamefont {Vallini}, \citenamefont {Amili},
  \citenamefont {Fainman},\ and\ \citenamefont {Kant{\'e}}}]{bahari2017b}%
  \BibitemOpen
  \bibfield  {author} {\bibinfo {author} {\bibfnamefont {B.}~\bibnamefont
  {Bahari}}, \bibinfo {author} {\bibfnamefont {A.}~\bibnamefont {Ndao}},
  \bibinfo {author} {\bibfnamefont {F.}~\bibnamefont {Vallini}}, \bibinfo
  {author} {\bibfnamefont {A.~E.}\ \bibnamefont {Amili}}, \bibinfo {author}
  {\bibfnamefont {Y.}~\bibnamefont {Fainman}}, \ and\ \bibinfo {author}
  {\bibfnamefont {B.}~\bibnamefont {Kant{\'e}}},\ }\href {\doibase
  10.1126/science.aao4551} {\bibfield  {journal} {\bibinfo  {journal}
  {Science}\ }\textbf {\bibinfo {volume} {358}},\ \bibinfo {pages} {636}
  (\bibinfo {year} {2017})}\BibitemShut {NoStop}%
\bibitem [{\citenamefont {Bandres}\ \emph {et~al.}(2018)\citenamefont
  {Bandres}, \citenamefont {Wittek}, \citenamefont {Harari}, \citenamefont
  {Parto}, \citenamefont {Ren}, \citenamefont {Segev}, \citenamefont
  {Christodoulides},\ and\ \citenamefont {Khajavikhan}}]{bandres2018}%
  \BibitemOpen
  \bibfield  {author} {\bibinfo {author} {\bibfnamefont {M.~A.}\ \bibnamefont
  {Bandres}}, \bibinfo {author} {\bibfnamefont {S.}~\bibnamefont {Wittek}},
  \bibinfo {author} {\bibfnamefont {G.}~\bibnamefont {Harari}}, \bibinfo
  {author} {\bibfnamefont {M.}~\bibnamefont {Parto}}, \bibinfo {author}
  {\bibfnamefont {J.}~\bibnamefont {Ren}}, \bibinfo {author} {\bibfnamefont
  {M.}~\bibnamefont {Segev}}, \bibinfo {author} {\bibfnamefont {D.~N.}\
  \bibnamefont {Christodoulides}}, \ and\ \bibinfo {author} {\bibfnamefont
  {M.}~\bibnamefont {Khajavikhan}},\ }\href {\doibase 10.1126/science.aar4005}
  {\bibfield  {journal} {\bibinfo  {journal} {Science}\ }\textbf {\bibinfo
  {volume} {359}},\ \bibinfo {pages} {eaar4005} (\bibinfo {year}
  {2018})}\BibitemShut {NoStop}%
\bibitem [{\citenamefont {Klembt}\ \emph {et~al.}(2018)\citenamefont {Klembt},
  \citenamefont {Harder}, \citenamefont {Egorov}, \citenamefont {Winkler},
  \citenamefont {Ge}, \citenamefont {Bandres}, \citenamefont {Emmerling},
  \citenamefont {Worschech}, \citenamefont {Liew}, \citenamefont {Segev},
  \citenamefont {Schneider},\ and\ \citenamefont {H{\"o}fling}}]{klembt2018}%
  \BibitemOpen
  \bibfield  {author} {\bibinfo {author} {\bibfnamefont {S.}~\bibnamefont
  {Klembt}}, \bibinfo {author} {\bibfnamefont {T.~H.}\ \bibnamefont {Harder}},
  \bibinfo {author} {\bibfnamefont {O.~A.}\ \bibnamefont {Egorov}}, \bibinfo
  {author} {\bibfnamefont {K.}~\bibnamefont {Winkler}}, \bibinfo {author}
  {\bibfnamefont {R.}~\bibnamefont {Ge}}, \bibinfo {author} {\bibfnamefont
  {M.~A.}\ \bibnamefont {Bandres}}, \bibinfo {author} {\bibfnamefont
  {M.}~\bibnamefont {Emmerling}}, \bibinfo {author} {\bibfnamefont
  {L.}~\bibnamefont {Worschech}}, \bibinfo {author} {\bibfnamefont {T.~C.~H.}\
  \bibnamefont {Liew}}, \bibinfo {author} {\bibfnamefont {M.}~\bibnamefont
  {Segev}}, \bibinfo {author} {\bibfnamefont {C.}~\bibnamefont {Schneider}}, \
  and\ \bibinfo {author} {\bibfnamefont {S.}~\bibnamefont {H{\"o}fling}},\
  }\href {\doibase 10.1038/s41586-018-0601-5} {\bibfield  {journal} {\bibinfo
  {journal} {Nature}\ }\textbf {\bibinfo {volume} {562}},\ \bibinfo {pages}
  {552} (\bibinfo {year} {2018})}\BibitemShut {NoStop}%
\bibitem [{\citenamefont {Zeng}\ \emph {et~al.}(2020)\citenamefont {Zeng},
  \citenamefont {Chattopadhyay}, \citenamefont {Zhu}, \citenamefont {Qiang},
  \citenamefont {Li}, \citenamefont {Jin}, \citenamefont {Li}, \citenamefont
  {Davies}, \citenamefont {Linfield}, \citenamefont {Zhang}, \citenamefont
  {Chong},\ and\ \citenamefont {Wang}}]{zeng2020}%
  \BibitemOpen
  \bibfield  {author} {\bibinfo {author} {\bibfnamefont {Y.}~\bibnamefont
  {Zeng}}, \bibinfo {author} {\bibfnamefont {U.}~\bibnamefont {Chattopadhyay}},
  \bibinfo {author} {\bibfnamefont {B.}~\bibnamefont {Zhu}}, \bibinfo {author}
  {\bibfnamefont {B.}~\bibnamefont {Qiang}}, \bibinfo {author} {\bibfnamefont
  {J.}~\bibnamefont {Li}}, \bibinfo {author} {\bibfnamefont {Y.}~\bibnamefont
  {Jin}}, \bibinfo {author} {\bibfnamefont {L.}~\bibnamefont {Li}}, \bibinfo
  {author} {\bibfnamefont {A.~G.}\ \bibnamefont {Davies}}, \bibinfo {author}
  {\bibfnamefont {E.~H.}\ \bibnamefont {Linfield}}, \bibinfo {author}
  {\bibfnamefont {B.}~\bibnamefont {Zhang}}, \bibinfo {author} {\bibfnamefont
  {Y.}~\bibnamefont {Chong}}, \ and\ \bibinfo {author} {\bibfnamefont {Q.~J.}\
  \bibnamefont {Wang}},\ }\href {\doibase 10.1038/s41586-020-1981-x} {\bibfield
   {journal} {\bibinfo  {journal} {Nature}\ }\textbf {\bibinfo {volume}
  {578}},\ \bibinfo {pages} {246} (\bibinfo {year} {2020})}\BibitemShut
  {NoStop}%
\bibitem [{\citenamefont {Harari}\ \emph {et~al.}(2018)\citenamefont {Harari},
  \citenamefont {Bandres}, \citenamefont {Lumer}, \citenamefont {Rechtsman},
  \citenamefont {Chong}, \citenamefont {Khajavikhan}, \citenamefont
  {Christodoulides},\ and\ \citenamefont {Segev}}]{harari2018}%
  \BibitemOpen
  \bibfield  {author} {\bibinfo {author} {\bibfnamefont {G.}~\bibnamefont
  {Harari}}, \bibinfo {author} {\bibfnamefont {M.~A.}\ \bibnamefont {Bandres}},
  \bibinfo {author} {\bibfnamefont {Y.}~\bibnamefont {Lumer}}, \bibinfo
  {author} {\bibfnamefont {M.~C.}\ \bibnamefont {Rechtsman}}, \bibinfo {author}
  {\bibfnamefont {Y.~D.}\ \bibnamefont {Chong}}, \bibinfo {author}
  {\bibfnamefont {M.}~\bibnamefont {Khajavikhan}}, \bibinfo {author}
  {\bibfnamefont {D.~N.}\ \bibnamefont {Christodoulides}}, \ and\ \bibinfo
  {author} {\bibfnamefont {M.}~\bibnamefont {Segev}},\ }\href {\doibase
  10.1126/science.aar4003} {\bibfield  {journal} {\bibinfo  {journal}
  {Science}\ }\textbf {\bibinfo {volume} {359}},\ \bibinfo {pages} {eaar4003}
  (\bibinfo {year} {2018})}\BibitemShut {NoStop}%
\bibitem [{\citenamefont {Longhi}\ \emph {et~al.}(2018)\citenamefont {Longhi},
  \citenamefont {Kominis},\ and\ \citenamefont {Kovanis}}]{longhi2018}%
  \BibitemOpen
  \bibfield  {author} {\bibinfo {author} {\bibfnamefont {S.}~\bibnamefont
  {Longhi}}, \bibinfo {author} {\bibfnamefont {Y.}~\bibnamefont {Kominis}}, \
  and\ \bibinfo {author} {\bibfnamefont {V.}~\bibnamefont {Kovanis}},\ }\href
  {\doibase 10.1209/0295-5075/122/14004} {\bibfield  {journal} {\bibinfo
  {journal} {EPL}\ }\textbf {\bibinfo {volume} {122}},\ \bibinfo {pages}
  {14004} (\bibinfo {year} {2018})}\BibitemShut {NoStop}%
\bibitem [{\citenamefont {Secl{\`i}}\ \emph {et~al.}(2019)\citenamefont
  {Secl{\`i}}, \citenamefont {Capone},\ and\ \citenamefont
  {Carusotto}}]{secli2019}%
  \BibitemOpen
  \bibfield  {author} {\bibinfo {author} {\bibfnamefont {M.}~\bibnamefont
  {Secl{\`i}}}, \bibinfo {author} {\bibfnamefont {M.}~\bibnamefont {Capone}}, \
  and\ \bibinfo {author} {\bibfnamefont {I.}~\bibnamefont {Carusotto}},\ }\href
  {\doibase 10.1103/PhysRevResearch.1.033148} {\bibfield  {journal} {\bibinfo
  {journal} {Phys. Rev. Research}\ }\textbf {\bibinfo {volume} {1}},\ \bibinfo
  {pages} {033148} (\bibinfo {year} {2019})}\BibitemShut {NoStop}%
\bibitem [{\citenamefont {{H. Haken}}(1985)}]{haken1985}%
  \BibitemOpen
  \bibfield  {author} {\bibinfo {author} {\bibnamefont {{H. Haken}}},\
  }\href@noop {} {\emph {\bibinfo {title} {Light ({{Volume}} 2)}}}\ (\bibinfo
  {publisher} {{North-Holland Physics Publishing}},\ \bibinfo {address}
  {{Amsterdam}},\ \bibinfo {year} {1985})\BibitemShut {NoStop}%
\bibitem [{\citenamefont {{C. W. Gardiner}}\ and\ \citenamefont
  {Zoller}(2004)}]{gardiner2004}%
  \BibitemOpen
  \bibfield  {author} {\bibinfo {author} {\bibnamefont {{C. W. Gardiner}}}\
  and\ \bibinfo {author} {\bibfnamefont {P.}~\bibnamefont {Zoller}},\
  }\href@noop {} {\emph {\bibinfo {title} {Quantum {{Noise}}}}},\ Springer
  {{Series}} in {{Synergetics}}\ (\bibinfo  {publisher} {{Springer}},\ \bibinfo
  {address} {{Berlin/Heidelberg}},\ \bibinfo {year} {2004})\BibitemShut
  {NoStop}%
\bibitem [{\citenamefont {Schawlow}\ and\ \citenamefont
  {Townes}(1958)}]{schawlow1958}%
  \BibitemOpen
  \bibfield  {author} {\bibinfo {author} {\bibfnamefont {A.~L.}\ \bibnamefont
  {Schawlow}}\ and\ \bibinfo {author} {\bibfnamefont {C.~H.}\ \bibnamefont
  {Townes}},\ }\href {\doibase 10.1103/PhysRev.112.1940} {\bibfield  {journal}
  {\bibinfo  {journal} {Phys. Rev.}\ }\textbf {\bibinfo {volume} {112}},\
  \bibinfo {pages} {1940} (\bibinfo {year} {1958})}\BibitemShut {NoStop}%
\bibitem [{\citenamefont {Pick}\ \emph {et~al.}(2015)\citenamefont {Pick},
  \citenamefont {Cerjan}, \citenamefont {Liu}, \citenamefont {Rodriguez},
  \citenamefont {Stone}, \citenamefont {Chong},\ and\ \citenamefont
  {Johnson}}]{pick2015}%
  \BibitemOpen
  \bibfield  {author} {\bibinfo {author} {\bibfnamefont {A.}~\bibnamefont
  {Pick}}, \bibinfo {author} {\bibfnamefont {A.}~\bibnamefont {Cerjan}},
  \bibinfo {author} {\bibfnamefont {D.}~\bibnamefont {Liu}}, \bibinfo {author}
  {\bibfnamefont {A.~W.}\ \bibnamefont {Rodriguez}}, \bibinfo {author}
  {\bibfnamefont {A.~D.}\ \bibnamefont {Stone}}, \bibinfo {author}
  {\bibfnamefont {Y.~D.}\ \bibnamefont {Chong}}, \ and\ \bibinfo {author}
  {\bibfnamefont {S.~G.}\ \bibnamefont {Johnson}},\ }\href {\doibase
  10.1103/PhysRevA.91.063806} {\bibfield  {journal} {\bibinfo  {journal} {Phys.
  Rev. A}\ }\textbf {\bibinfo {volume} {91}},\ \bibinfo {pages} {063806}
  (\bibinfo {year} {2015})}\BibitemShut {NoStop}%
\bibitem [{\citenamefont {Hafezi}\ \emph {et~al.}(2013)\citenamefont {Hafezi},
  \citenamefont {Mittal}, \citenamefont {Fan}, \citenamefont {Migdall},\ and\
  \citenamefont {Taylor}}]{hafezi2013}%
  \BibitemOpen
  \bibfield  {author} {\bibinfo {author} {\bibfnamefont {M.}~\bibnamefont
  {Hafezi}}, \bibinfo {author} {\bibfnamefont {S.}~\bibnamefont {Mittal}},
  \bibinfo {author} {\bibfnamefont {J.}~\bibnamefont {Fan}}, \bibinfo {author}
  {\bibfnamefont {A.}~\bibnamefont {Migdall}}, \ and\ \bibinfo {author}
  {\bibfnamefont {J.~M.}\ \bibnamefont {Taylor}},\ }\href {\doibase
  10.1038/nphoton.2013.274} {\bibfield  {journal} {\bibinfo  {journal} {Nat.
  Photonics}\ }\textbf {\bibinfo {volume} {7}},\ \bibinfo {pages} {1001}
  (\bibinfo {year} {2013})}\BibitemShut {NoStop}%
\bibitem [{\citenamefont {Haldane}(1988)}]{haldane1988}%
  \BibitemOpen
  \bibfield  {author} {\bibinfo {author} {\bibfnamefont {F.~D.~M.}\
  \bibnamefont {Haldane}},\ }\href {\doibase 10.1103/PhysRevLett.61.2015}
  {\bibfield  {journal} {\bibinfo  {journal} {Phys. Rev. Lett.}\ }\textbf
  {\bibinfo {volume} {61}},\ \bibinfo {pages} {2015} (\bibinfo {year}
  {1988})}\BibitemShut {NoStop}%
\bibitem [{\citenamefont {Jotzu}\ \emph {et~al.}(2014)\citenamefont {Jotzu},
  \citenamefont {Messer}, \citenamefont {Desbuquois}, \citenamefont {Lebrat},
  \citenamefont {Uehlinger}, \citenamefont {Greif},\ and\ \citenamefont
  {Esslinger}}]{jotzu2014}%
  \BibitemOpen
  \bibfield  {author} {\bibinfo {author} {\bibfnamefont {G.}~\bibnamefont
  {Jotzu}}, \bibinfo {author} {\bibfnamefont {M.}~\bibnamefont {Messer}},
  \bibinfo {author} {\bibfnamefont {R.}~\bibnamefont {Desbuquois}}, \bibinfo
  {author} {\bibfnamefont {M.}~\bibnamefont {Lebrat}}, \bibinfo {author}
  {\bibfnamefont {T.}~\bibnamefont {Uehlinger}}, \bibinfo {author}
  {\bibfnamefont {D.}~\bibnamefont {Greif}}, \ and\ \bibinfo {author}
  {\bibfnamefont {T.}~\bibnamefont {Esslinger}},\ }\href {\doibase
  10.1038/nature13915} {\bibfield  {journal} {\bibinfo  {journal} {Nature}\
  }\textbf {\bibinfo {volume} {515}},\ \bibinfo {pages} {237} (\bibinfo {year}
  {2014})}\BibitemShut {NoStop}%
\bibitem [{\citenamefont {Savvidis}\ \emph {et~al.}(2001)\citenamefont
  {Savvidis}, \citenamefont {Ciuti}, \citenamefont {Baumberg}, \citenamefont
  {Whittaker}, \citenamefont {Skolnick},\ and\ \citenamefont
  {Roberts}}]{savvidis2001}%
  \BibitemOpen
  \bibfield  {author} {\bibinfo {author} {\bibfnamefont {P.~G.}\ \bibnamefont
  {Savvidis}}, \bibinfo {author} {\bibfnamefont {C.}~\bibnamefont {Ciuti}},
  \bibinfo {author} {\bibfnamefont {J.~J.}\ \bibnamefont {Baumberg}}, \bibinfo
  {author} {\bibfnamefont {D.~M.}\ \bibnamefont {Whittaker}}, \bibinfo {author}
  {\bibfnamefont {M.~S.}\ \bibnamefont {Skolnick}}, \ and\ \bibinfo {author}
  {\bibfnamefont {J.~S.}\ \bibnamefont {Roberts}},\ }\href {\doibase
  10.1103/PhysRevB.64.075311} {\bibfield  {journal} {\bibinfo  {journal} {Phys.
  Rev. B}\ }\textbf {\bibinfo {volume} {64}},\ \bibinfo {pages} {075311}
  (\bibinfo {year} {2001})}\BibitemShut {NoStop}%
\bibitem [{\citenamefont {Bernier}\ \emph {et~al.}(2014)\citenamefont
  {Bernier}, \citenamefont {Dalla~Torre},\ and\ \citenamefont
  {Demler}}]{bernier2014}%
  \BibitemOpen
  \bibfield  {author} {\bibinfo {author} {\bibfnamefont {N.~R.}\ \bibnamefont
  {Bernier}}, \bibinfo {author} {\bibfnamefont {E.~G.}\ \bibnamefont
  {Dalla~Torre}}, \ and\ \bibinfo {author} {\bibfnamefont {E.}~\bibnamefont
  {Demler}},\ }\href {\doibase 10.1103/PhysRevLett.113.065303} {\bibfield
  {journal} {\bibinfo  {journal} {Phys. Rev. Lett.}\ }\textbf {\bibinfo
  {volume} {113}},\ \bibinfo {pages} {065303} (\bibinfo {year}
  {2014})}\BibitemShut {NoStop}%
\bibitem [{\citenamefont {Bernier}\ \emph {et~al.}(2018)\citenamefont
  {Bernier}, \citenamefont {T{\'o}th}, \citenamefont {Feofanov},\ and\
  \citenamefont {Kippenberg}}]{bernier2018}%
  \BibitemOpen
  \bibfield  {author} {\bibinfo {author} {\bibfnamefont {N.~R.}\ \bibnamefont
  {Bernier}}, \bibinfo {author} {\bibfnamefont {L.~D.}\ \bibnamefont
  {T{\'o}th}}, \bibinfo {author} {\bibfnamefont {A.~K.}\ \bibnamefont
  {Feofanov}}, \ and\ \bibinfo {author} {\bibfnamefont {T.~J.}\ \bibnamefont
  {Kippenberg}},\ }\href {\doibase 10.1103/PhysRevA.98.023841} {\bibfield
  {journal} {\bibinfo  {journal} {Phys. Rev. A}\ }\textbf {\bibinfo {volume}
  {98}},\ \bibinfo {pages} {023841} (\bibinfo {year} {2018})}\BibitemShut
  {NoStop}%
\bibitem [{\citenamefont {Hodaei}\ \emph {et~al.}(2014)\citenamefont {Hodaei},
  \citenamefont {Miri}, \citenamefont {Heinrich}, \citenamefont
  {Christodoulides},\ and\ \citenamefont {Khajavikhan}}]{hodaei2014}%
  \BibitemOpen
  \bibfield  {author} {\bibinfo {author} {\bibfnamefont {H.}~\bibnamefont
  {Hodaei}}, \bibinfo {author} {\bibfnamefont {M.-A.}\ \bibnamefont {Miri}},
  \bibinfo {author} {\bibfnamefont {M.}~\bibnamefont {Heinrich}}, \bibinfo
  {author} {\bibfnamefont {D.~N.}\ \bibnamefont {Christodoulides}}, \ and\
  \bibinfo {author} {\bibfnamefont {M.}~\bibnamefont {Khajavikhan}},\ }\href
  {\doibase 10.1126/science.1258480} {\bibfield  {journal} {\bibinfo  {journal}
  {Science}\ }\textbf {\bibinfo {volume} {346}},\ \bibinfo {pages} {975}
  (\bibinfo {year} {2014})}\BibitemShut {NoStop}%
\bibitem [{\citenamefont {Hafezi}\ \emph {et~al.}(2011)\citenamefont {Hafezi},
  \citenamefont {Demler}, \citenamefont {Lukin},\ and\ \citenamefont
  {Taylor}}]{hafezi2011}%
  \BibitemOpen
  \bibfield  {author} {\bibinfo {author} {\bibfnamefont {M.}~\bibnamefont
  {Hafezi}}, \bibinfo {author} {\bibfnamefont {E.~A.}\ \bibnamefont {Demler}},
  \bibinfo {author} {\bibfnamefont {M.~D.}\ \bibnamefont {Lukin}}, \ and\
  \bibinfo {author} {\bibfnamefont {J.~M.}\ \bibnamefont {Taylor}},\ }\href
  {\doibase 10.1038/nphys2063} {\bibfield  {journal} {\bibinfo  {journal} {Nat.
  Phys.}\ }\textbf {\bibinfo {volume} {7}},\ \bibinfo {pages} {907} (\bibinfo
  {year} {2011})}\BibitemShut {NoStop}%
\bibitem [{\citenamefont {Amelio}\ and\ \citenamefont
  {Carusotto}(2019)}]{amelio2019}%
  \BibitemOpen
  \bibfield  {author} {\bibinfo {author} {\bibfnamefont {I.}~\bibnamefont
  {Amelio}}\ and\ \bibinfo {author} {\bibfnamefont {I.}~\bibnamefont
  {Carusotto}},\ }\href@noop {} {\bibfield  {journal} {\bibinfo  {journal}
  {arXiv:1911.10437}\ } (\bibinfo {year} {2019})}\BibitemShut {NoStop}%
\end{thebibliography}%

\end{document}